# A Review on Environmental Barrier Coatings: History, Current State of the Art and Future Developments

Daniel Tejero-Martin, Chris Bennett, Tanvir Hussain*

*Faculty of Engineering, University of Nottingham, Nottingham, NG7 2RD, UK*

* +44 115 951 3795, tanvir.hussain@nottingham.ac.uk

**Abstract**

The increasing demand for more efficient and environmental-friendly gas turbines has driven the development of new strategies for material development. SiC/SiC ceramic matrix composites (CMCs) can fulfil the stringent requirements; however, they require protection from the operating environment and debris ingested during operation. Environmental barrier coatings (EBCs) are a protective measure to enable the CMCs to operate under harsh conditions. EBC-coated CMCs will enable an increased efficiency and reduced pollutant and $CO_2$ emissions. In this review, the fundamentals of SiC/SiC ceramic matrix composites degradation in steam environments and under the presence of molten alkali salts, namely $CaO-MgO-Al_2O_3-SiO_2$ (CMAS), are first presented. Then, a summary of EBCs along with a comprehensive summary of the current compositions and their interactions with steam and molten salts is presented. Finally, an overview of the latest research directions for the potential next generation of EBCs are outlined.

# 1. Introduction

Gas turbine engines for aerospace and energy generation represent the cornerstone of a rapidly growing sector, with an estimation of 2 trillion USD for the cumulative sales of gas turbine engines in the 2017-2031 period [1]. Due to this considerable economic presence, there is a great interest for the development of better performing components. Ni-based super-alloys have been for the past decades the norm for components in the hot section of gas turbine engines. Improvements on thermal barrier coatings and cooling mechanisms have allowed the industry to increase the gas inlet temperatures up to 1500 °C [2,3], driving upwards the thermal efficiency, the thrust-to-weight ratio and reducing the emission of noxious by-products. Nevertheless, this strategy is approaching the intrinsic limit imposed by the melting point of Ni-based super-alloys, and novel strategies will be needed to further increase the gas inlet temperature. A new approach is required for the next breakthrough in jet engines, and SiC/SiC ceramic matrix composites are the most promising material to fulfil the role. When compared to Ni-based super-alloys, CMCs provide an increased melting point and superior strength at high temperature, as it can be seen on Figure 1.

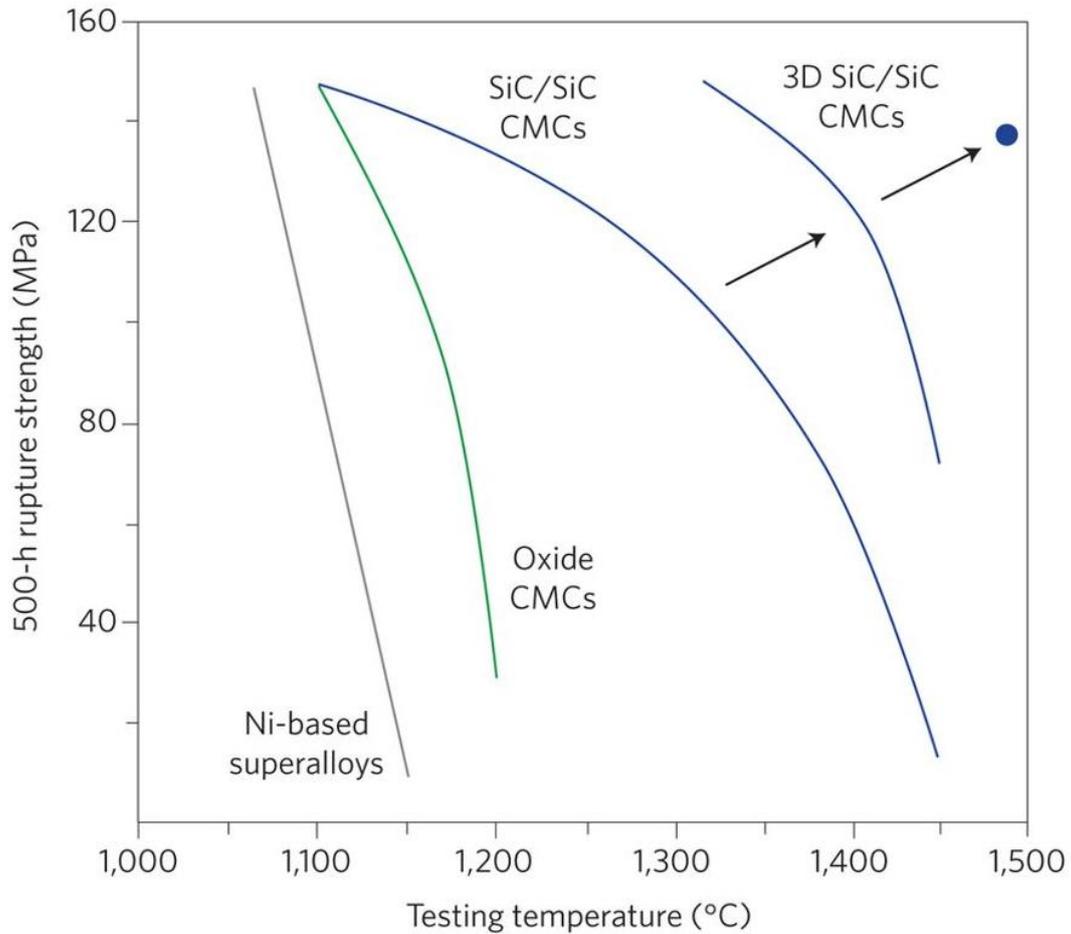

*Figure 1: Rupture strength after 500 h of continuous exposure versus testing temperature of Ni-based superalloys, oxide CMCs and various SiC/SiC CMCs. The blue point is the 300 h rupture strength* [2]

Under clean, dry oxygen atmosphere, SiC-based CMCs present excellent oxidation resistance attributed to the formation of a protective silica layer. Al Nasiri *et al.* [4] reported the oxidation kinetics of SiC/SiC CMCs exposed at 1200 – 1400 °C for up to 48 h in air, showing a parabolic behaviour of the oxidation reaction kinetics leading to a decrease in the oxidation rate associated with the diffusion of oxygen through the oxide layer, with an activation energy of 619 kJ/mol. Nevertheless, under the presence of steam (a common combustion reaction product) or molten alkali salts and reducing environments (caused by the ingestion of debris with the intake air or present as fuel impurities) accelerated degradation of the otherwise protective silica layer takes place, compromising the integrity of the CMC [5–9]. The effect of steam on CMCs has been extensively studied over the past decades, since it was realised early in their development that an increase in the steam content led to an accelerated oxidation rate. In the late twentieth century Opila *et al.* [5] first studied the precise mechanism behind silica volatilisation, establishing an additional step following the reaction of SiC with $O_2$ to form silica, shown in reaction (1) in which the silica further reacted with $H_2O$ to form gaseous Si-O-H species, such as $Si(OH)_4$ [6] as shown in reaction (2), therefore causing the mentioned silica volatilisation.

$$SiC + 1.5O_2(g) = SiO_2 + CO(g) \qquad (1)$$

$$SiO_2 + 2H_2O(g) = Si(OH)_4(g) \qquad (2)$$

This volatilisation process would be accompanied by a recession of the surface of the component, which has been calculated to be as high as ~1 µm/h under normal gas turbine operating conditions (temperature of 1350 °C, gas velocity of 300 m/s, steam partial pressure of 0.1 atm and total pressure of 1 atm) [10]. Such recession rate would imply an unacceptable level of corrosion for components that are expected to operate without maintenance for at least 30,000 h.

On the other hand, the negative effect of molten salts has also been extensively studied for decades. Contrary to the case of steam, salt degradation can be caused by a wide range of chemical compounds, making its study and prevention more challenging. An early NASA report from the late 1980s [11] focused on the research being conducted since the 1970s on SiC degradation caused by $Na_2SO_4$ on heat engines, caused by the operation of jet engines over marine environments, or due to Na impurities present in the fuel. The report clearly reinforces the fact that, despite the potential breakthrough that SiC components could represent, protective measurements would be first required. A first attempt to limit the corrosion experienced by SiC was reported by Federer [12] with the application of several alumina-based coatings. In his work, thermal cycling up to 1200 °C and corrosion testing at 1200 °C in a $Na_2CO_4$ containing atmosphere were used, with the results indicating that mullite ($3Al_2O_3 2SiO_2$) provided the best match of coefficient of thermal expansion (CTE), preventing spallation due to stresses during thermal cycling, and improved resistance under corrosive conditions. Along with the discovery of mullite as a promising candidate in the protection of SiC components, these experiments also remarked the importance of a closely matched coefficient of thermal expansion between SiC and the deposited coatings, a recurring challenge in the development of protective systems. In the early 1990s reports were coming in regarding the effect of the ingestion of volcanic material by planes flying near volcanic plumes [13]. This promoted a shift from the previously studied hot corrosion by $Na_2SO_4$ from a more generalised family of compositions, being labelled as $CaO-MgO-Al_2O_3-SiO_2$ or CMAS [14,15]. Therefore, as summarised above, the early realisation of the deleterious effect of steam and corrosive compounds (such as alkali salts or debris) on the longevity of CMCs prompted the desire to develop a protective coating that would prevent the environmental attack of SiC components. With this goal in mind, environmental barrier coatings were first introduced as a solution to the exacerbated corrosion experienced by CMCs under typical service environments. As it has been mentioned above, EBCs are expected to fulfil a set of requirements in order to be considered fit for service, being the five main characteristics shown in Figure 2.

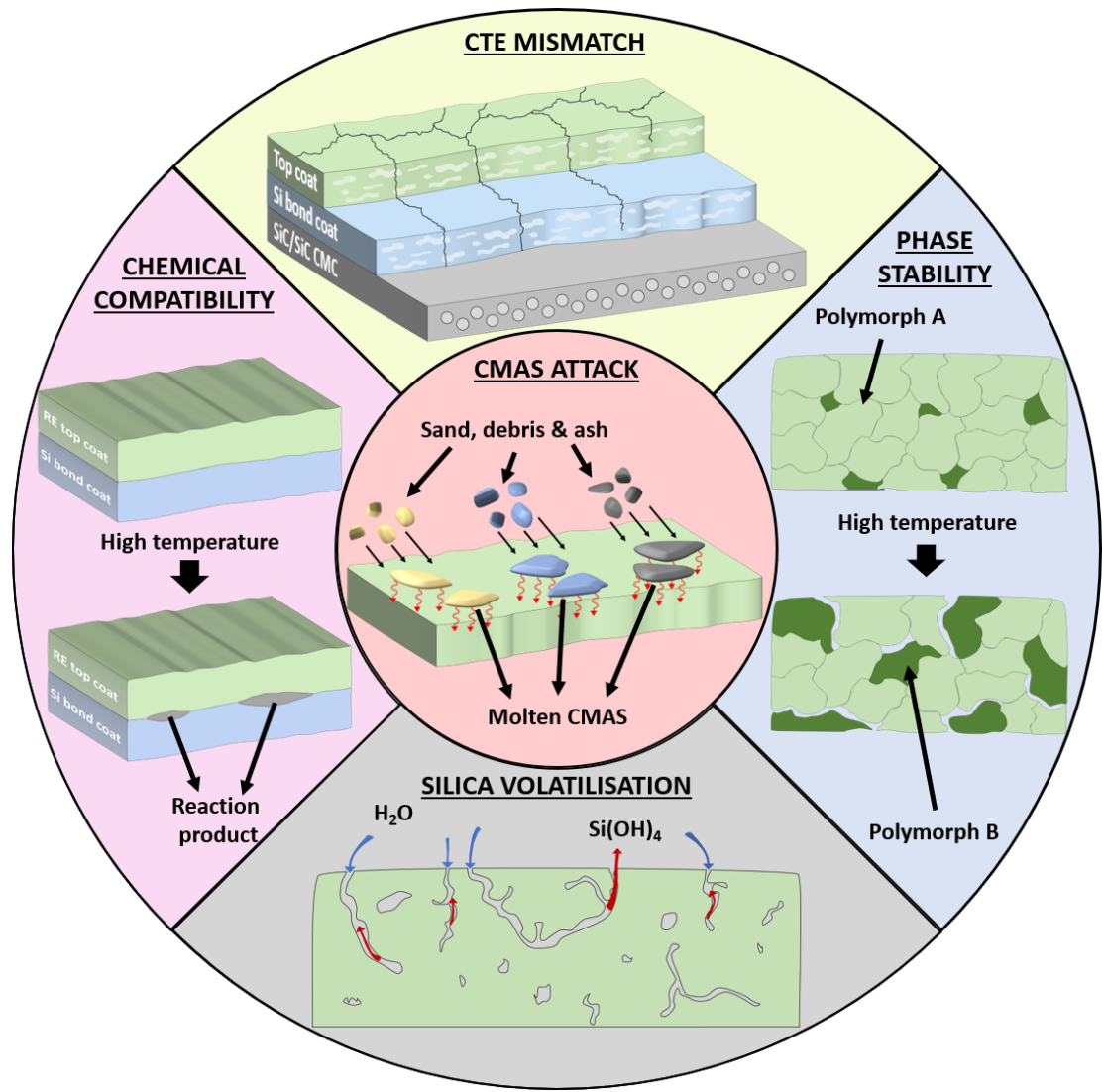

*Figure 2: Schematic of the main five requirements that any successful EBC is expected to fulfil.*

As it can be seen in Figure 2 high temperature induces a series of phenomena that determine whether the EBC will remain protective and fulfil its role, or fail. At the top, a representation of the effects of a mismatch in the CTE is shown. Due to the presence of heating and cooling cycles during service, thermal expansion and contraction will take place for each of the components of the EBC. If the magnitude of their CTE is too different from each other, the thermal stresses induced will lead to the formation of cracks. On the right sector it can be seen how the presence of polymorphs can affect. The as-deposited coating might show a majority of polymorph A, but at high temperatures there might be a phase transformation into polymorph B, which can be accompanied by a noticeable volume contraction (or expansion), causing cracking and defects such as porosity. At the bottom, the process of silica volatilisation is presented. The presence of steam at high temperature induces the formation of silica containing gases, such as $Si(OH)_4$, producing the recession of the material. On the left, the chemical compatibility between the layers present is shown. Materials that at room temperature might show good compatibility and stability might become reactive and produce unwanted by-products when exposed to high temperatures for extended periods of time. Finally, in the middle, the presence of various debris and impurities leads to molten deposits of salts (generalised under the term CMAS, $CaO$-$MgO$-$Al_2O_3$-$SiO_2$) that can have detrimental effects on the coating. As it can be seen, the development of a successful EBC is a complex task that has required, and still requires to this day, extensive research. As with any challenge, many unsuccessful approaches have been tried for the field to move forward. This never-ending search for more optimised solutions is represented in Figure 3, where a timeline with the evolution of the most notable compositions for EBCs is shown.

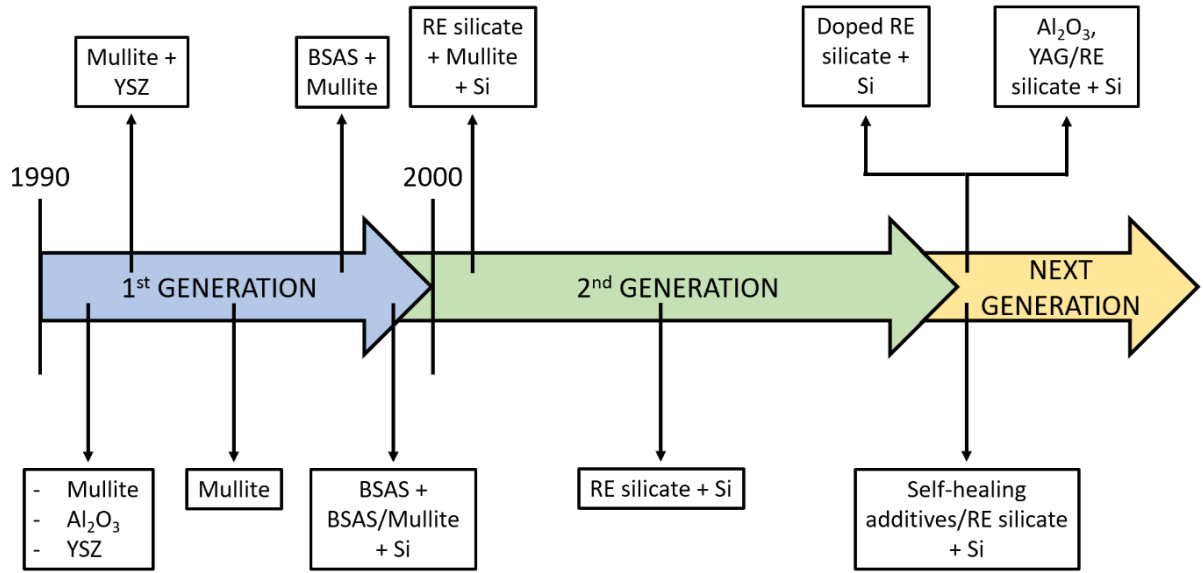

*Figure 3: Timeline of the evolution in the design of EBCs, including some of the major compositions used*

In this work, a historical overview of the development of EBCs is first presented, with a special focus on the process that established the actual requirement for state-of-the-art EBC compositions. A detailed review of the current most promising candidates is presented, referring to the specifications mentioned above, and their behaviour under the most common environments (steam oxidation and CMAS corrosion). Finally, an overview of some of the future developments in the field of EBCs is summarised.

## 2. Development of environmental barrier coatings

As it was shown in Figure 3, the development of EBCs is generally categorised into different generations based on the main composition being used. In this section, a more detailed review of the different generations is presented, with a particular focus on the capabilities and disadvantages that led to next compositions, and the knowledge gathered.

### 2.1. First generation

The first generation of EBCs is usually delimited by the initial developments in the 1990s, involving mullite and BSAS, up to the introduction of rare earth silicates around the beginning of the 2000s. A more detailed description of each composition is presented in the following sections.

#### 2.1.1. Mullite

Initial works were based on the discoveries made by Federer [12], with mullite being the prime candidate investigated due to its environmental durability, chemical compatibility and CTE match with SiC, with SiC having a value of ~4.5 × $10^{-6}$ °C$^{-1}$ and mullite being ~5 × $10^{-6}$ °C$^{-1}$ [16]. Further research was conducted on the protective capabilities of air plasma sprayed (APS) refractory oxide coatings, such as mullite, yttria-stabilised zirconia (YSZ), alumina ($Al_2O_3$) and a mixture of them [17]. It was proven that mullite had the capability to stay attached to the sprayed components while providing protection against corrosive environments. Despite the already mentioned CTE match, APS mullite coatings presented cracks after thermal cycling, severely compromising its ability to protect the substrate. Research on the phase stability and microstructure of the coatings showed that the cause of the failure was not CTE mismatch, but the crystallisation of residual amounts of metastable amorphous mullite formed due to the rapid cooling during APS deposition [18]. When exposed to temperatures above 1000 °C, mullite crystallises, a process that involves a volume change causing cracking, as shown in Figure 4. In order to prevent this, fully crystalline mullite was deposited while maintaining the substrate above the crystallisation temperature, reducing the appearance of cracks after ten 20 h cycles up to 1400 °C and showing promising results after exposure to $Na_2CO_3$ at 1000 °C for 24 h. The investigation of mullite failure led to another requirement for any design of successful EBCs: along with a CTE close to that of SiC, the coating must maintain a stable phase under thermal exposure.

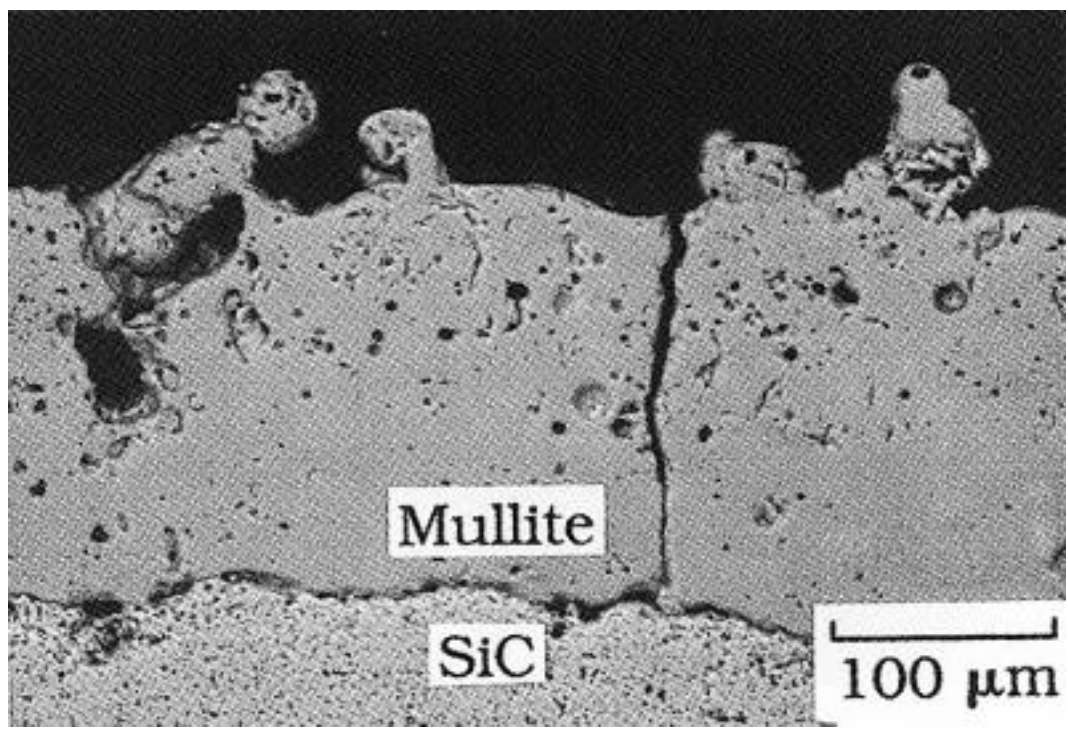

*Figure 4: Mullite coated SiC after ten 20 h cycles between room temperature and 1400 °C showing cracks due to the crystallisation process* [18]

Making use of the improved deposition methodology, it was shown that plasma-sprayed mullite coatings were capable of withstanding thermal exposure at 1300 °C in air for 1200 h [19] and high pressure hot corrosion burner rig testing at 1000 °C for 150 h [20]. Nevertheless, by the end of the 1990s, research on silica volatilisation in the presence of steam for SiC components, along with the new finding that indicated a high silica activity (~0.4) for mullite under similar conditions [21], shifted research efforts from molten salts towards steam-resistant EBCs.

An initial approach to minimise the silica volatilisation experienced by the mullite coatings was, inspired by its success as thermal barrier coatings, the use of yttria-stabilised zirconia (YSZ) as an additional top coat. In this configuration, the mullite layer was applied as a bond coat, with the aim of not only providing oxidation/corrosion resistance to the SiC substrate, but also allowing the bonding of the YSZ ($ZrO_2$ – 8wt.% $Y_2O_3$) overlayer [20,22,23]. Despite the great success of YSZ as a thermal barrier coating, the large CTE mismatch between mullite (5 × $10^{-6}$ °C) and YSZ (~10 × $10^{-6}$ °C) coupled with residual phase transformation in the plasma-sprayed mullite bond coat, severely limited the durability of the EBC. Due to the formation of thermal stresses, cracking was induced during cyclic thermal exposure, and a preferential pathway for the ingress of steam was created. For these reasons, YSZ was promptly discarded and new compositions were tested.

### 2.1.2. BSAS

BSAS (1-xBaO·xSrO·$Al_2O_3$·$2SiO_2$, 0 ≤ x ≤ 1), a new composition, was proposed as part of the high speed research-enabling propulsion materials (HSR-EPM) program [24], being derived from the well-studied mullite. A better matched CTE (~4.5 × $10^{-6}$ °C for the monoclinic celsian phase [25]) and a lower silica activity than mullite (<0.1) [26] produced a more crack-resistant coating, which improved the EBC service time. The durability of the EBC system was further improved through the modification of the mullite bond coat via the addition of BSAS, forming a mullite-BSAS composite bond coat leading to a reduced presence of cracks. Finally, the multi-layered EBC system was further improved through the addition of a silicon layer at the mullite-BSAS bond coat/SiC interface, effectively increasing the adherence [25,27]. By the end of the 1990s and beginning of the 2000s, the state-of-the-art EBC was formed of three layers: a silicon bond coat, a mullite-BSAS intermediate coat and a BSAS top coat. This design was proven on SiC/SiC CMC combustor liners for three Solar Turbine industrial gas turbine engines, with a total operation time of over 24,000 h without failure, and a Texaco engine successfully completing a 14,000 h test [16,28].

Despite the clear advancements made possible by this first generation of EBCs, some issues still limited the performance of the coatings and the maximum temperature capable of withstanding. When BSAS was first identified as a promising EBC, due to the fulfilment of the two main requirements described so far, namely CTE match with SiC and phase stability, tests were conducted to evaluate the performance of BSAS directly deposited on CMC. Thermal cycling at 1300 °C under an atmosphere of 90% $H_2O$ – 10% $O_2$ for 100 h showed extensive reaction between BSAS and the thermally grown silica layer, causing large pores at the interface [29]. The reaction caused the appearance of a low-melting (~1300 °C) glass product and interfacial porosity. Such pores were formed due to the bubbling of gaseous species, product of the BSAS – silica reaction, ultimately leading to the spallation of the coating. In order to overcome the nefarious interaction between BSAS and silica, a chemical barrier was applied. The already mentioned modified mullite-BSAS bond coat provided an improvement on the durability of the EBC, showing reduced oxidation when tested under the same conditions described above. The addition of a bond coat did not, however, completely prevent the reaction between BSAS and silica. After 1000 h at 1300 °C under an atmosphere of 90% $H_2O$ – 10% $O_2$, evidence of this glassy by-product was observed, as shown in Figure 5.

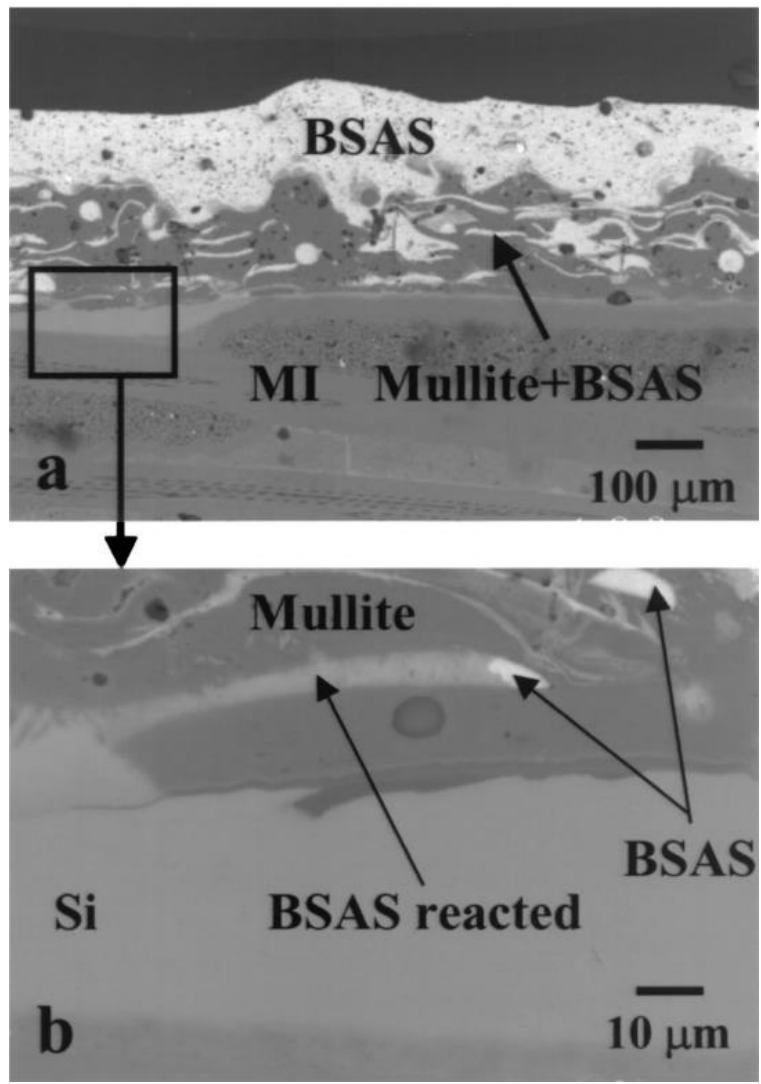

*Figure 5: Cross section of the Si/(mullite-BSAS)/BSAS EBC deposited on melt infiltrated SiC/SiC continuous-fibre-reinforced composite (MI CFC) after 100 h in 90% $H_2O$ - 10% $O_2$ at 1300 °C [29]*

The issue was only aggravated with increasing temperature, with the glassy product penetrating through the BSAS top coat and showing on the surface of the EBC after 300 h at 1400 °C in 90% $H_2O$ – 10% $O_2$. This imposed an upper limit of 1300 °C for the application of BSAS-based EBC solutions, effectively negating the intended purpose of EBCs: allowing an increase in the gas inlet temperature beyond 1400 °C. The replacement of BSAS as the main component for EBCs was needed, although it allowed the recognition of a new essential requirement for the design of future EBCs: high temperature chemical compatibility between layers.

## 2.2. Second generation

Despite the great advancements achieved since the first iteration of EBCs, it was soon realised that mullite and BSAS based systems would not live up to the expectations. As mentioned previously, the path towards a new composition was made possible by the knowledge gained during the initial experimental and theoretical work. To summarise the requirements identified at the end of the 1990s, a successful EBC system had to present the following characteristics. Firstly, a close CTE match between the forming layers and the SiC substrate is required to avoid thermal stresses and the appearance of cracks. Next, it is expected that the coating does not undergo any phase transformations during high temperature exposure, or at least, that if a phase transformation does occur, the CTE of the involved polymorphs is close in value and there is minimal volume change. Thirdly, an EBC must be characterised by a low silica activity under a variety of conditions, such as dry or wet environments. Finally, highlighted by the experimental evidence that a multi-layered systems would be required, a chemical compatibility must exist between the involved compositions of the different layers, in order to avoid the formation of unwanted and detrimental reaction products at the interfaces, risking the structural integrity of the EBC and altering its protective capabilities.

Therefore, once it was realised that mullite and BSAS were not ideal candidates for the ambitious goals in mind, and with a clear set of requirements for the next generation of EBCs, a research program was launched at NASA in 1999. The initiative, named the ultraefficient engine technology (UEET) programme, had the goal to conduct extensive research and screening tests to identify the prime materials capable of withstanding a temperature of 1316 °C (2400 °F) at the EBC – SiC substrate interface and 1482 °C (2700 °F) at the EBC surface for thousands of hours [30]. This programme identified a new family of compositions with promising properties, being categorised under the name of rare earth silicates. Within rare earth silicates, two main compositions are present, namely rare earth monosilicates ($RE_2SiO_5$, being RE a rare earth element) and rare earth disilicates ($RE_2Si_2O_7$). Among the rare earth silicates identified as suitable candidates, were those with rare elements such as scandium (Sc), lutetium (Lu), ytterbium (Yb), yttrium (Y) and erbium (Er) [16].

As mentioned before, the first condition that any potential composition has to fulfil in order to be considered for its use as EBC is a close CTE match with the SiC substrate. Table 1 shows the CTE of a selected range of rare earth silicates along with that of SiC and silicon, in addition to the space group categorisation according to the Felsche classification in the case of rare earth silicates [31].

| Composition | Space group | Average CTE ( x $10^{-6}$ $K^{-1}$) | Reference |
| --- | --- | --- | --- |
| SiC | | 4.5 – 5.5 | [32] |
| Si | | 3.5 – 4.5 | [32] |
| β - $Sc_2Si_2O_7$ | $C2/m$ | 5.4 | [33] |
| $Lu_2SiO_5$ | $I2/a$ | 6.7 | [34] |
| β - $Lu_2Si_2O_7$ | $C2/m$ | 4.2 | [33] |
| $Yb_2SiO_5$ | $I2/a$ | 7.1 – 7.4 | [34] |
| $Yb_2SiO_5$ | $P2_1/c$ | --- | |
| β - $Yb_2Si_2O_7$ | $C2/m$ | 3.6 – 4.5 | [33] |
| X1 - $Y_2SiO_5$ | $P2_1/c$ | 8.7 | [34] |
| X2 - $Y_2SiO_5$ | $I2/a$ | 6 – 7.7 | [34] |
| α - $Y_2Si_2O_7$ | $P\bar{1}$ | 8 | [35] |
| β - $Y_2Si_2O_7$ | $C2/m$ | 3.6 – 4.5 | [33,35] |
| γ - $Y_2Si_2O_7$ | $P2_1/c$ | 3.9 | [35] |
| δ - Y$_2$Si$_2$O$_7$ | $Pnam$ | 8.1 | [35] |
| Er$_2$SiO$_5$ | $I2/a$ | 5 - 7 | [34] |
| β - Er$_2$Si$_2$O$_7$ | $C2/m$ | 3.9 | [33] |

*Table 1: Space group and average CTE for several rare earth silicates considered for its use as EBCs*

A closely matched CTE is not the only requirement for an EBC, and as it can be seen in Table 1, several rare earth silicates present polymorphs. This will produce a phase transformation at high temperatures, as shown in Figure 6, which in most cases is undesirable due to a potential abrupt change in the CTE.

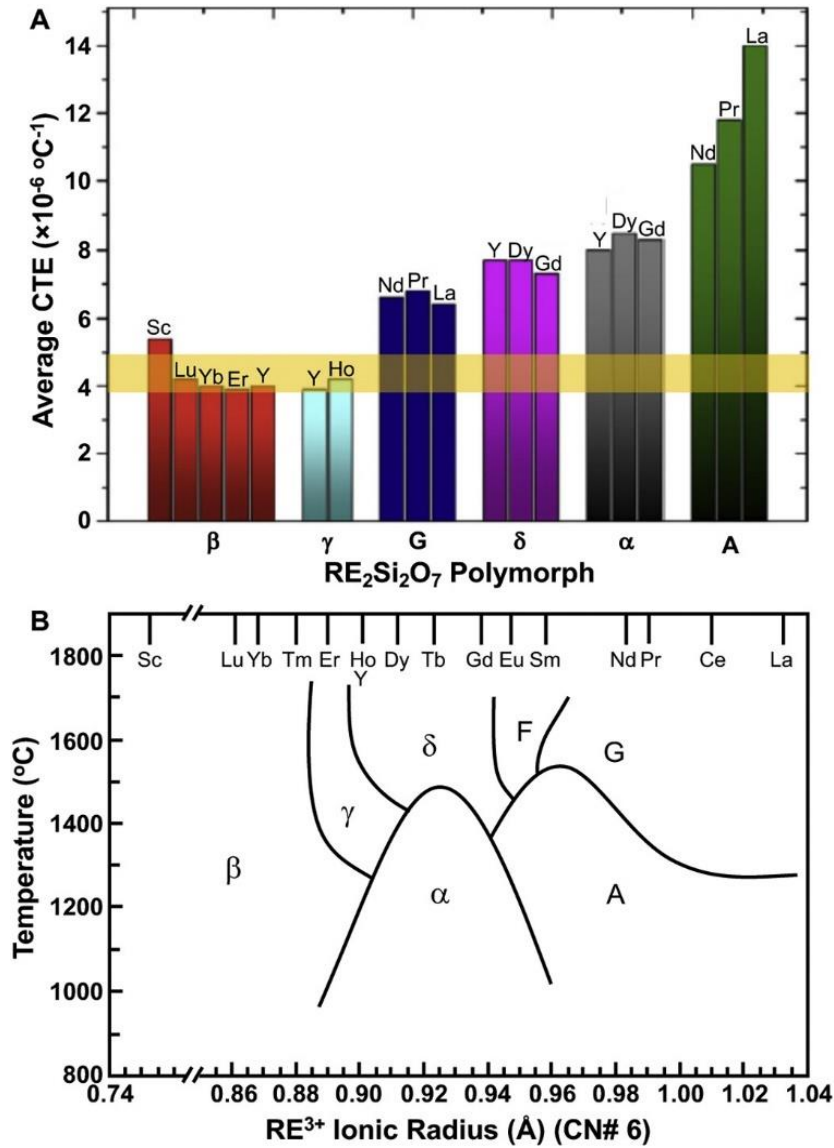

*Figure 6: (A) Average CTE of rare earth disilicates polymorphs. The horizontal band indicates the range of CTE values for SiC CMCs. (B) Diagram of the different polymorphs present within rare earth disilicates according to temperature* [36]

When studying the silica activity of rare earth silicates, it is assumed that only Si(OH)$_4$ is removed as a gaseous sub-product, being the rest of the EBC components rapidly disintegrated [37]. In that case the volatilisation rate can be linked to the removal of SiO$_2$ from the rare earth silicate in the form of gaseous

Si(OH)$_4$, as it was shown in Equation 2. The total recession suffered by the system is then directly proportional to the silica activity $a_{SiO_2}$ of the otherwise protective top coat. The higher the silica activity, the higher the level of volatilisation, which eventually will lead to unacceptable recession levels on components that are expected to provide protection for up to 30,000 h. This relationship is described in equations (4) and (5) below, where the weight loss rates (*k*) for silica volatilisation in the case of laminar and turbulent flow conditions are shown:

$$k_{laminar} = a_{SiO_2} \cdot exp\left(-\frac{E}{RT}\right) \cdot v^{1/2} \cdot \left(P_{H_2O}\right)^n \cdot P^{-1/2} \tag{4}$$

$$k_{turbulent} = a_{SiO_2} \cdot exp\left(-\frac{E}{RT}\right) \cdot v^{4/5} \cdot \left(P_{H_2O}\right)^n \cdot P^{-1/5} \tag{5}$$

Where *E* is the activation energy, *R* is the gas constant, *T* is the temperature, *v* is the gas velocity, $P_{H_2O}$ is the steam partial pressure, *n* is the steam partial pressure exponent and *P* is the total pressure. Due to this direct connection between the silica activity and the recession rate, a reliable database containing the $a_{SiO_2}$ values of the main EBC candidates would be an invaluable tool; however, testing and measuring this is not a trivial task. As it can be seen, the flow conditions of the gas (laminar vs turbulent), the flow velocity, steam partial pressure and total pressure also play a role, which makes reliably measuring $a_{SiO_2}$ quite a challenge. Measurements performed at specific test conditions might not be entirely comparable to others performed under different conditions, and lab-based testing systems might differ greatly from the expected conditions during service, as shown in Figure 7 for the recession rate of SiC. In addition to the intrinsic problematic nature of the task, external considerations such as the material of the furnace tube should be taken into account. If fused quartz (SiO$_2$) tubes are used, the hot steam will corrode the tube, artificially increasing the level of Si(OH)$_4$ experienced by the samples. Whereas, if alumina (Al$_2$O$_3$) tubes are used, contamination will be produced through Al(OH)$_3$, promoting the formation of compositions otherwise not expected [1,16,38].

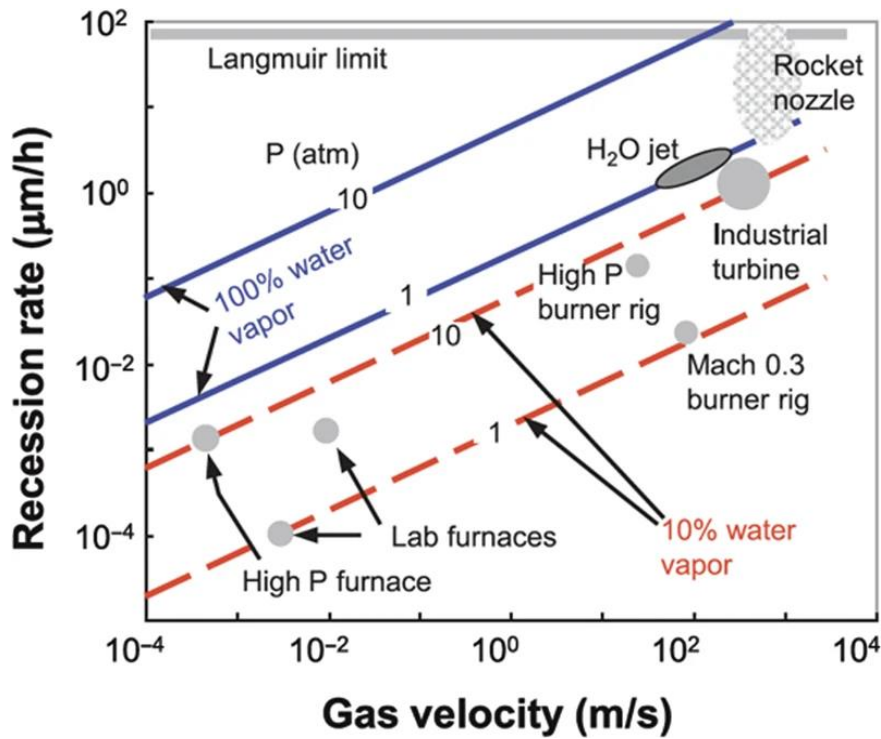

*Figure 7: Recession rate for SiC under different conditions, in all cases being the temperature 1316 °C and assuming linear flow [1]*

Despite these challenges, considerable effort (both experimental and computational) has been put into developing reliable testing methodologies to assess the volatility of different materials under high temperature, high velocity steam flow [10,39,40]. Providing specific values for the silica volatility of rare earth silicates may not yet be possible, as theoretical calculations and experimental measurements still differ too much. Worth mentioning is the work of Jacobson [41] and Costa and Jacobson [42] in the calculation of the theoretical values for the $Y_2O_3$-$SiO_2$ and $Yb_2O_3$-$SiO_2$ systems, showing good agreement with some experimental values measured in the lab for YDS-YMS and YbDS-YbMS coatings, respectively. Nevertheless, there is still too much variability in the volatilisation rate of rare earth silicates to definitively validate the theoretical calculations. As a general conclusion, however, it is agreed that the volatility of monosilicates is lower than its disilicate counterpart, measured under identical conditions, although the CTE of the latter tends to be closer to that of SiC and Si. Finally, no evidence of chemical incompatibility between rare earth silicates and the rest of the layers commonly applied in an EBC system has been reported, which seems to indicate that this family of materials is an ideal candidate for its use as an EBC.

### 2.2.1. Lutetium silicates

Lutetium silicates have been studied as they lack polymorphs and their initial volatility measurements provided promising results. Extensive work has been conducted by Ueno *et al.* [43–47] aiming to determine the recession rate of both Lu mono- and disilicate under a variety of steam testing conditions. Their studies concluded that although Lu silicates have good characteristic to become successful EBCs, its performance under steam testing conditions was less than ideal. During the synthesis process of the disilicate, an incorrect ratio between the precursor compositions causes the formation of both LuDS and LuMS, with the addition of $SiO_2$ located in the grain boundary, as shown in Equation 6.

$$Lu_2O_3(s) + 2SiO_2(s) = (1-x)Lu_2Si_2O_7(s) + xLu_2SiO_5(s) + SiO_2(boundary) \qquad (6)$$

The presence of silica at the grain boundary was the cause of the failure of the coatings, since its removal upon interaction with the steam caused the formation of gas paths within the coating, as can be seen in Figure 8. These gas paths provided an entry for oxidisers into the substrate underneath, causing excessive oxidation and mass gain of the tested samples.

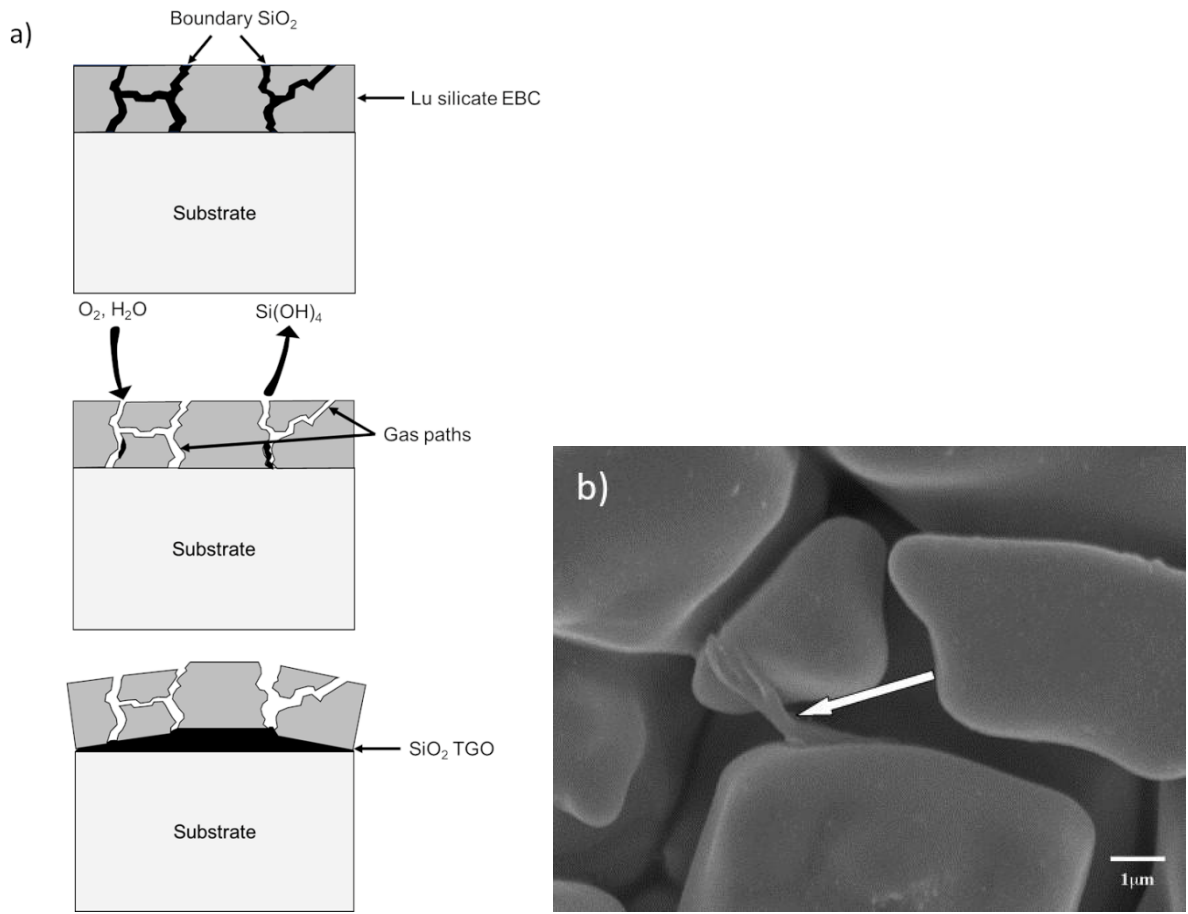

*Figure 8: a) Schematic of the corrosion attack caused by hot steam on a Lu silicate EBC layer on a silicon nitride substrate, showing the volatilisation of the $SiO_2$ grain boundary and the access of oxidisers to the substrate underneath. b) SEM image of the corrosion experienced at the grain boundaries after steam testing at 1300 °C with 30%$H_2O$/70% air for 100 h and a gas flow rate of 175 ml/min. Redrawn from [44,46]*

Nevertheless, additional reports of Lu silicate coatings exposed to steam testing [38,48,49] show that perhaps a suitable deposition method could be applied to avoid the severe grain boundary corrosion, making these compositions worthy of additional research to clarify whether they can be successfully applied as EBCs or not. Despite these advancements, lutetium does not currently stand as a preferential candidate for a successful EBC, mainly due to the presence of other rare earth elements that show fewer issues when exposed to steam and with similar volatility rates (i.e. yttrium and ytterbium). As it stands, lutetium silicates are an interesting option for research into potential deposition methods that lead to satisfactory steam performance, but not for its industrial application.

### 2.2.2. Yttrium silicates

Yttrium silicates were among the first rare earth silicates to be studied, with a brief mention in the literature that a yttrium silicate coating deposited using thermal spray had been tested for 500 h of cyclic steam testing at 1200 °C showing [28], although limited details were disclosed due to patents in the same system being granted [50]. Although the presence of multiple polymorphs for both YMS and YDS, as shown in Table 1, could be a limiting factor for the application of Y-based EBCs, some authors have reported that the γ-$Y_2Si_2O_7$ polymorph is stable between ~1320 °C and temperature above 1600 °C, showing a sluggish transformation kinetics down to 1200 °C [38,51]. Regarding the two YMS polymorphs, X1-YMS and X2-YMS, there have been reports of transformation between the low temperature X1 to the high temperature X2 polymorph after exposure to 100% steam to 1100 °C and 1200 °C for up to 16 h with gas flow of 2 $m^3$/h and gas velocity of 5 m/s [52]; however, exposure to 1300 °C caused the coating to decompose, with only $Y_2O_3$ being detected, which represents a clear failure for a material expected to withstand thousands of hours at high temperatures. No such decomposition has been reported by other authors, although the high CTE value of YMS (around 8 × $10^{-6}$ $K^{-1}$ for both polymorphs, compared to ~5 × $10^{-6}$ $K^{-1}$ for SiC) and a slightly higher silica activity when compared to other rare earth silicates [38] and the presence of a multitude of polymorphs for YDS has made yttrium silicate lose momentum against ytterbium silicates when it comes to potential candidates for EBC applications.

In summary, yttrium silicates have some attractive properties that could have made them ideal candidates for EBC systems; however, they add additional degrees of complexity to the design of a protective multilayer solution. The presence of several polymorphs with a wide range of CTE values in the case of disilicates, despite the reports of sluggish phase transformation and potential high temperature stability, represents another variable that should be taken into account and prevented during operation. In the case of monosilicates, contradictory reports exist regarding its high temperature stability, but phase transformation between its two polymorphs seems well documented, which is certainly a disadvantage. Seeing that the general trend in the design of EBC is the simplification of the system (with the removal of the mullite layer, favouring just a rare earth silicate top coat and Si bond coat) makes the choice of yttrium silicates quite difficult to argue for.

### 2.2.3. Ytterbium silicates

Initial efforts were focused on ytterbium monosilicate (YbMS - $Yb_2SiO_5$) as a promising EBC candidate due to its lower silica volatility. The research carried by Richards *et al.* [53,54] on the deposition of rare earth silicates top coat + mullite diffusion barrier + Si bond coat using plasma spraying (APS) and the study of the characteristics of the deposited coatings before and after thermal cycling in water environment, provided great insights into the failure mechanisms [55,56]. Drawing from the knowledge on the mullite transformation at high temperatures, both the mullite diffusion layer and the rare earth silicate top coat were deposited using atmospheric air plasma with the substrate heated to 1200 °C inside a box furnace. Despite the use of this improved deposition procedure, the defects and porosities present in the coating, combined with the CTE mismatch between the layers, produced the failure of the system through the appearance of vertical cracks (mud cracks), as shown in Figure 9, after annealing in air at 1300 °C for 20 h. Those same mud cracks were partially responsible for the poor performance of the coating under steam cycling conditions (1 atm pressure, flow velocity of 4.4 cm/s, 90% $H_2O$/10% $O_2$ environment, 60 min at 1316 °C and 10 min at 110 °C), presenting spallation after less than 200 cycles. It was concluded that despite the considerable efforts concerning the optimisation of the deposition method, the characteristics of the deposited coating and the adherence between the layers, the CTE mismatch was too great to produce a successful EBC. Vertical cracks would appear during the heat treatments, as it can be seen in Figure 9, providing a preferential path for the ingress of oxidisers, which then reacted with the Si bond coat, producing the failure of the EBC.

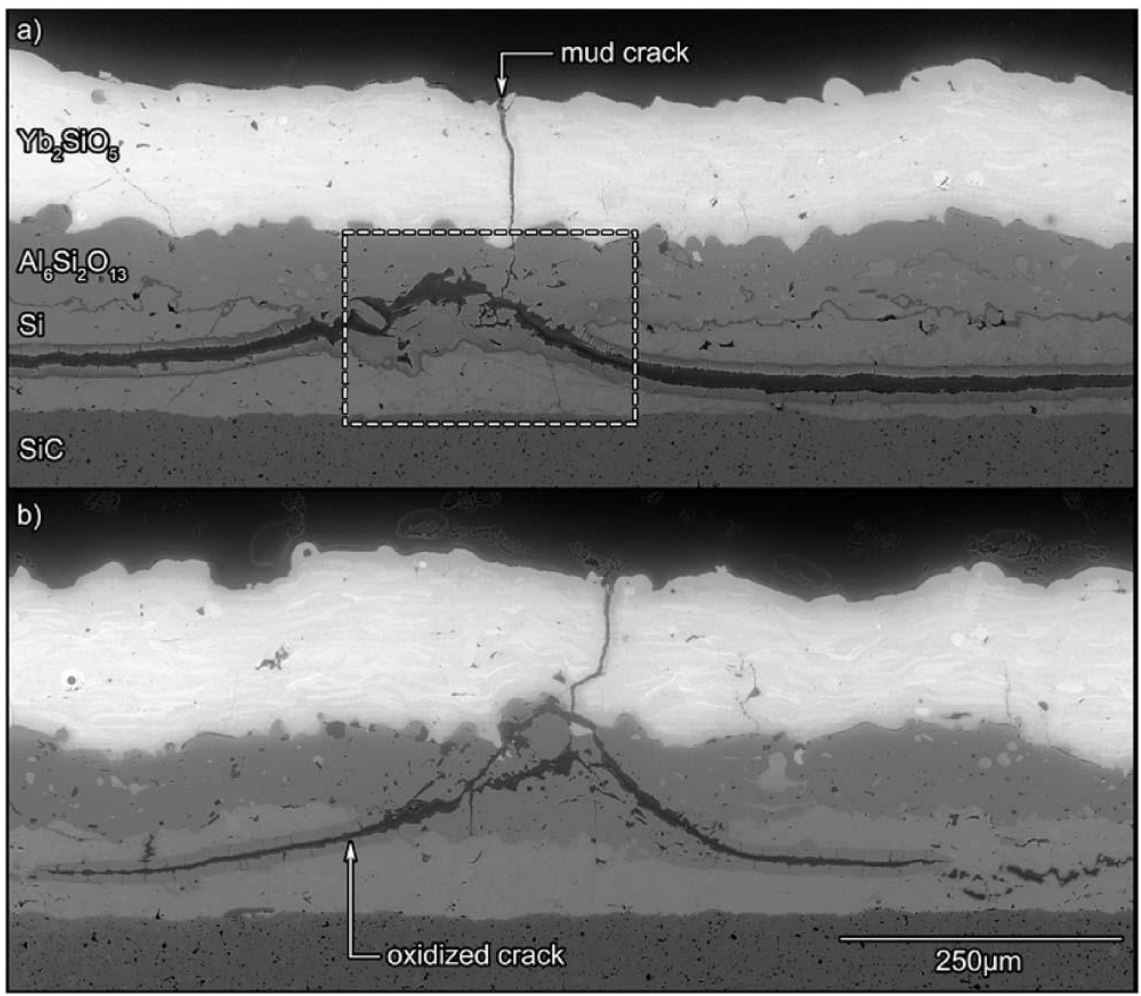

*Figure 9: SEM images of the cross section of the failed coating after 250 1-h steam cycles* [56]

Seeing that CTE match still represents one of the main challenges when selecting a potential material, the attention was shifted towards rare earth disilicates, which despite having a slightly higher silica activity, tend to present a better matched CTE with the SiC substrate. Ytterbium disilicate (YbDS – $Yb_2Si_2O_7$) has been primarily studied as a potential candidate for its use as an EBC. The preferential volatilisation of YbDS versus YbMS was also found by Bakan *et al.* [57] when studying the oxidation behaviour under steam conditions (1 atm pressure, flow velocity of 100 m/s, partial pressure of steam of 0.15 atm, temperature of 1200 °C and up to 200 h of exposure) of APS deposited coatings with different YbDS/YbMS ratio. The coating deposited at higher plasma power resulted in a lower YbDS content (36 wt.%) due to the increased $SiO_2$ loss during spraying, whereas a lower plasma power produced a coating with a higher YbDS (62 wt.%). The YbDS content was directly correlated to the mass loss and severity of the corrosion during the test, showing better performance in the case of low YbDS content. Nevertheless, phase content is not the only factor that should be considered. Microstructure, particularly porosity and pore connectivity also play an essential role. Bakan et al. [58] reported the corrosion of APS deposited coatings with YbDS content of 70 wt.% compared to sintered bodies with YbDS content of 92 wt.%. Their results show that weight loss and the thickness of the $SiO_2$ depleted layer was smaller in the case of the sintering body, due to the reduced porosity level and lower pore connectivity.

Therefore, the focus of this section will be in thermal sprayed coatings, as they better represent the characteristics and performance of the final product. Initial reports showed that the deposition of YbDS using thermal spray could yield low-porosity, mud-crack-free coatings [54], providing an exciting new candidate. Further investigations on the deposition of YbDS using thermal spray techniques have been conducted [59,60], being worth noting the work of Garcia et al. [61] on the effect of different $SiO_2$ content in the feedstock powder and plasma power when using APS to achieve the desired composition. The different rates of volatilisation of $SiO_2$ affect the viscosity of the splats and phase composition of the coating, modifying the initial state of the as sprayed coatings. Their work highlighted the importance that crystallisation and phase transformation have in the integrity of the coatings. Crystallisation, since the APS deposited coatings will present an amorphous state in the as sprayed condition, due to the rapid cooling involved in the process [62], and will crystallise once heat treated. Phase transformation as continued exposure to high temperatures will promote the conversion of *P2$_1$/c* $Yb_2SiO_5$ into the *I2/a* polymorph, accompanied by a volume expansion. The combination of $SiO_2$ volatilisation during spraying (and the associated appearance of $Yb_2SiO_5$), volume contraction due to crystallisation, volume expansion due to phase transformation, thermal stresses arising from CTE mismatch, changes in porosity and formation of a thermal grown oxide at the top coat/Si bond coat interface deepens the complexity of designing a successful EBC system.

Despite the availability of several thermal spraying deposition techniques, efforts have been currently focused on the use of APS, although several techniques have been preliminarily considered. An example is the work of Bakan et al. [63,64] on the deposition of YbDS coatings using APS, suspension plasma spray (SPS), high velocity oxy-fuel (HVOF) spray and very-low pressure plasma (VLPPS) spray. Their work shows how the different melting levels of the splats achieved with each technique, as well as the subsequent cooling rate of the coated samples, affect the degree of crystallinity and the structural integrity of the coatings. High temperature techniques, such as APS and SPS, produced highly amorphous coatings that cracked during the post-deposition cooling to room temperature. HVOF sprayed samples presented a higher degree of crystallinity, due to the presence of semi-molten and non-molten particles as confirmed by electron backscatter diffraction [64], and higher porosity levels. The reduced thermal stresses related to a lower flame temperature, and the increased presence of porosity, key elements to a better strain tolerance, resulted in crack-free coatings. Regarding VLPPS, the ability to maintain the substrates heated to a temperature close to 1000 °C prior to spraying and the use of the plasma flame to reduce the post-deposition cooling rate gave rise to highly crystalline coatings with no visible cracks. Despite being VLPPS the technique that produced better results, its application for real sized components might not be achievable in terms of operations cost and size limitations of the vacuum chamber. Therefore, HVOF thermal spray might be an interesting alternative for future EBC developments.

Of all the rare earth silicate top coats presented in this work, ytterbium compositions seem to have currently the advantage in terms of favourable properties for its application by the industry. Further research is still needed, particularly in suitable deposition techniques that fulfil the requirements needed for a successful EBC while presenting a realistic technique applicable on a large scale on components of large dimensions and complex shapes. Although this preference for ytterbium silicates is based on its inherent properties and high temperature behaviour, there is still room for improvement, and future research into more complex compositions with ytterbium silicates as a base might represent the future of the field, as it is discussed in more detail in section "Next generation of EBC".

## 3. Corrosion mechanisms

During the development of better performing EBCs, steam and molten alkali salts were early identified as the main challenges presented in terms of corrosion encounter during service. A brief description of the deleterious effects of both corrodents have been presented in the previous sections; however, a more detailed review of the fundamentals for both steam and molten alkali salts is presented here. For the sake of brevity, this comprehensive summary will only cover the corrosion mechanisms and effects reported in rare earth silicate EBCs.

### 3.1. Steam degradation

When considering the effect that flowing steam at high temperature has on EBC systems it is necessary to remember that an EBC is expected to behave as a gas-tight layer, reducing the penetration of oxidisers to the substrate underneath. Nevertheless, as it has been shown before, several factors can affect the physical integrity of the EBC, in which case the coating loses its gas-tight characteristic, causing the system to experience a shortened effective life. Two main mechanisms can be identified as the cause for the structural failure of the coatings. First, as described before, flowing steam at elevated temperatures will induce the volatilisation of silica from the rare earth silicate top coat. Even in the ideal scenario of a homogenous material removal from the coating, this mass loss will eventually lead to the failure of the coating, leaving exposed the unprotected component beneath. On top of this, material removal due to silica volatilisation is rarely a homogenous process in coatings. Differences in the phase content, porosity level and surface roughness cause hot spots for volatilisation and erosion due to the flowing steam, producing accelerated material removal, as reported by Bakan *et al.* [57] and shown in Figure 10.

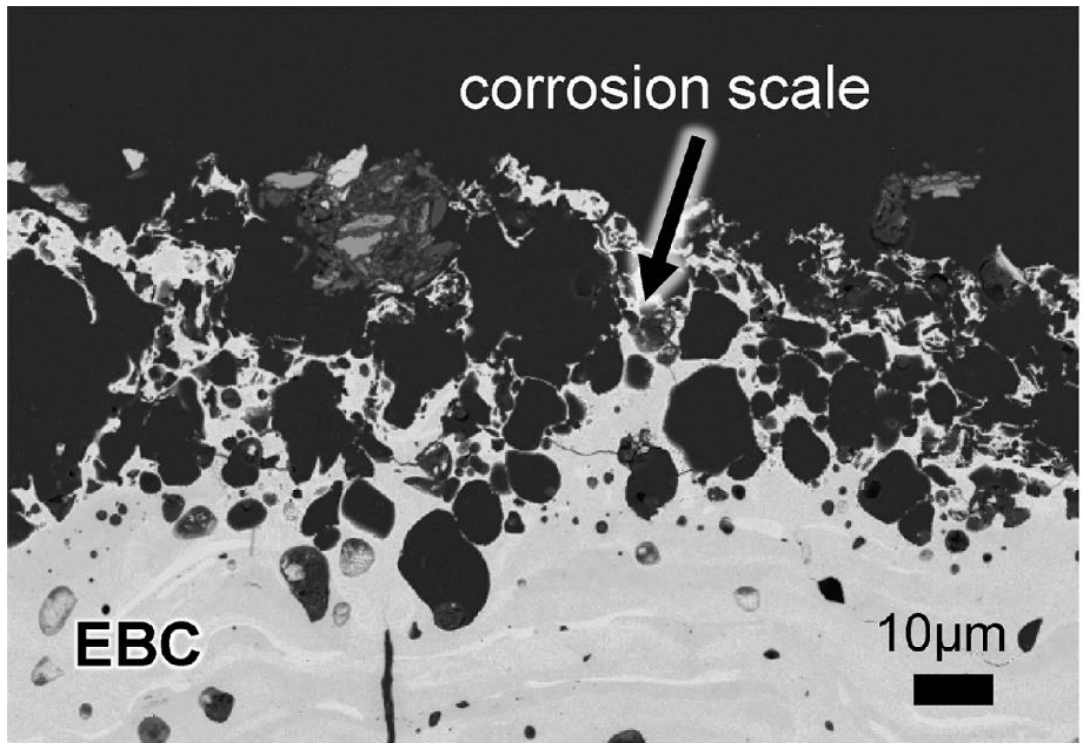

*Figure 10: BSE SEM image of the cross section of a ytterbium disilicate coating deposited using APS and exposed to steam corrosion (temperature: 1200 °C, gas flow velocity: 100 m/s, $P_{H_2O}$ = 0.15 atm, $P_{total}$ = 1 atm, time = 200 h)* [57]

For the purpose of comparison, Table 2 shows a summary of the corrosion under steam conditions of rare earth silicates measured so far, including the maximum mass loss, the volatility rate (if available) and the deposition method. A few details should be taken into consideration when consulting Table 2. First, the deposition method (including whether the test was performed on a single layer or on a coating deposited onto a substrate), the phase composition as deposited and after the test, and the porosity are provided (if reported) as this factors will greatly affect the volatilisation. Secondly, the maximum corrosion experienced (whether mass weight gained or lost) at the end of the testing time is indicated as a measure of the total damage experienced, although this value is hardly comparable between different reports for the reasons previously mentioned. A volatilisation rate is provided, which is a better value for comparison. This volatilisation rate is expressed as reported from the literature, calculated if the volatilisation at different time points was reported and linear behaviour was observed or estimated from the maximum corrosion if linear behaviour was observed.

| Material | Deposition method | As deposited phase composition | Post-testing phase composition | Porosity | Testing conditions | Max volatilisation | Volatilisation rate | Reference |
|---|---|---|---|---|---|---|---|---|
| $Y_2SiO_5$ (YMS) | Uniaxial cold pressing (50 MPa) + sintering at 1580 °C for 3 h | YMS: at least 85 wt.% | --- | 2 % | Temperature: 1350 °C. Flow rate: 40 ml/min. Composition: 90%$H_2O$/10%$O_2$. Testing up to 166 h | -0.404 mg/cm$^2$ after 166 h | -0.00258 mg/cm$^2$·h | [48] |
| | Hot pressing at 1500 °C/27.6 MPa in vacuum | YMS. Traces of $Y_2O_3$ and YDS | YMS. Traces of $Al_2Y_4O_9$ | --- | Temperature: 1500 °C. Flow velocity: 4.4 cm/s. Composition: 50%$H_2O$/10%$O_2$. Testing up to 100 h | 0.3 & 0.6 mg/cm$^2$ after 100 h | 0.003 & 0.006 mg/cm$^2$·h | [16] |
| | Magnetron sputtering on top of substrate + annealing at 1100 °C for 3 h in vacuum | X1- and X2-YMS polymorphs. Traces of $Y_{4.69}(SiO_4)_3O$ | $Y_2O_3$ | --- | Temperature: 1300 °C. Flow rate: 2 m$^3$/h. Flow velocity: 5 m/s. Composition: 100%$H_2O$. Testing up to 1 h | -1.22 mg/cm$^2$ after 1 h | --- | [52] |
| | | | X2-YMS and YDS | --- | Temperature: 1200 °C. Flow rate: 2 m$^3$/h. Flow velocity: 5 m/s. Composition: 100%$H_2O$. Testing up to 16 h | -1.18 mg/cm$^2$ after 16h | | |
| | Sol-gel + calcination at 1000 °C for 10 h | YMS | YMS, $Y_3Al_2(AlO_4)_3$, $Y_{4.67}(SiO_4)_3O$, YDS | --- | Temperature: 1400 °C. Flow velocity (cold zone): 5 cm/s. P$H_2O$ = 50 kPa, Ptotal = 100 kPa. Testing up to 300 - 310 h | 0.6 mg/cm$^2$ after 310 h | --- | [65] |
| | Sol-gel + calcination at 1000 °C for 10 h + sintering at 1400 °C for 5 h | | | --- | | 0.7 mg/cm$^2$ after 310 h | | |
| | Sol-gel + calcination at 1000 °C for 10 h + sintering at 1500 °C for 5 h | | | --- | | 0.45 mg/cm$^2$ after 310 h | | |
| | Milling and compacting powders, no sintering | $Y_2O_3$, $SiO_2$ | YMS, $Y_3Al_2(AlO_4)_3$, $Y_{4.67}(SiO_4)_3O$ | --- | | 1.45 mg/cm$^2$ after 300 h | | |
| | Milling and compacting powders + sintering at 1400 °C for 5 h | YMS, $Y_2O_3$ | | --- | | 0.6 mg/cm$^2$ after 310 h | | |
| | Milling and compacting powders + sintering at 1500 °C for 5 h | YMS | | --- | | 0.9 mg/cm$^2$ after 300 h | | |
| $Y_2Si_2O_7$ (YDS) | Cold pressing + sintering at 1500 °C for 2 h in Ar atmosphere | YDS, some YMS | YDS, decreased content of YMS | --- | Temperature: 1400 °C. Composition: 50%$H_2O$/50%$O_2$. Testing up to 400 h | -0.25 mg/cm$^2$ after 400 h | -0.00063 mg/cm$^2$·h | [66] |
| | Cold pressing + sintering at 1400 - 1600 °C | >99 wt.% YDS | --- | 30 % | Temperature: 1500 °C. Flow rate: ~290 l/h. Flow velocity: 13 cm/s. Composition 30%$H_2O$/70% air. P$H_2O$ = 0.3 bar, Ptotal = 0.1 MPa. Testing up to 310 h | -0.898 mg/cm$^2$ after 310 h | -0.00192 mg/cm$^2$·h | [38] |
| | Sol-gel + calcination at 1000 °C for 10 h | YDS | YDS | --- | Temperature: 1400 °C. Flow velocity (cold zone): 5 cm/s. P$H_2O$ = 50 kPa, Ptotal = 100 kPa. Testing up to 300 - 310 h | -2.5 mg/cm$^2$ after 310 h | --- | [65] |
| | Sol-gel + calcination at 1000 °C for 10 h + sintering at 1400 °C for 5 h | | | --- | | -1.8 mg/cm$^2$ after 310 h | | |
| | Sol-gel + calcination at 1000 °C for 10 h + sintering at 1500 °C for 5 h | | | --- | | -2.1 mg/cm$^2$ after 310 h | | |
| | Milling and compacting powders, no sintering | $Y_2O_3$, $SiO_2$ | YDS, $Y_3Al_2(AlO_4)_3$ | --- | | -0.1 mg/cm$^2$ after 300 h | | |
| | Milling and compacting powders + sintering at 1400 °C for 5 h | YDS, $Y_2O_3$ | | --- | | -0.5 mg/cm$^2$ after 310 h | | |
| | Milling and compacting powders + sintering at 1500 °C for 5 h | YDS | | --- | | -0.25 mg/cm$^2$ after 300 h | | |

| Material | Deposition method | As deposited phase composition | Post-testing phase composition | Porosity | Testing conditions | Max volatilisation | Volatilisation rate | Reference |
|---|---|---|---|---|---|---|---|---|
| $Gd_2SiO_5$ (GdMS) | Uniaxial cold pressing (50 MPa) + sintering at 1580 °C for 3 h | GdMS: 95 wt.% / GdDS: 5 wt.% | --- | 2 % | Temperature: 1350 °C. Flow rate: 40 ml/min. Composition: 90%$H_2O$/10%$O_2$. Testing up to 166 h | -2.3 mg/cm$^2$ after 166 h | -0.01576 mg/cm$^2$·h | [48] |
| $Er_2SiO_5$ (ErMS) | Uniaxial cold pressing (50 MPa) + sintering at 1580 °C for 12 h | ErMS: at least 85 wt.% | --- | 5 % | Temperature: 1350 °C. Flow rate: 40 ml/min. Composition: 90%$H_2O$/10%$O_2$. Testing up to 166 h | -0.502 mg/cm$^2$ after 166 h | -0.00353 mg/cm$^2$·h | [48] |
| | Hot pressing at 1500 °C/27.6 MPa in vacuum | ErMS. Traces of $Er_2O_3$ and ErDS | ErMS. Traces of $Al_{10}Er_6O_{24}$ | --- | Temperature: 1500 °C. Flow velocity: 4.4 cm/s. Composition: 50%$H_2O$/10%$O_2$. Testing up to 100 h | -0.1 mg/cm$^2$ after 100 h | Not linear | [16] |
| $Yb_2SiO_5$ (YbMS) | Uniaxial cold pressing (50 MPa) + sintering at 1580 °C for 3 h | YbMS: 85 wt.% / YbDS: 15 wt.% | --- | 6 % | Temperature: 1350 °C. Flow rate: 40 ml/min. Composition: 90%$H_2O$/10%$O_2$. Testing up to 166 h | -0.347 mg/cm$^2$ after 166 h | -0.00213 mg/cm$^2$·h | [48] |
| | Hot pressing at 1500 °C/27.6 MPa in vacuum | YbMS. Traces of $Yb_2O_3$ and YbDS | YbMS. Traces of $Al_5Yb_3O_{12}$ | --- | Temperature: 1500 °C. Flow velocity: 4.4 cm/s. Composition: 50%$H_2O$/10%$O_2$. Testing up to 100 h | 0.05 mg/cm$^2$ after 100 | Not linear | [16] |
| | Dip coating CMC substrate + heat treatment at 1350 °C for 50 h | > 90 wt.% YbMS, < 10 wt.% YbDS, $Yb_2O_3$ | Mainly YbDS | ~ 10 % | Temperature: 1350 °C. Flow rate: 0.67 cm$^3$/s. Composition: 90%$H_2O$/10%$O_2$. Testing up to 150 h | 0.55 mg/cm$^2$ after 150 h | 0.00277 mg/cm$^2$·h | [49] |
| $Yb_2Si_2O_7$ (YbDS) | Hot pressing at 1500 °C/27.6 MPa in vacuum | YbDS. Traces of YbMS | YbDS. Traces of YbMS and $Al_5Yb_3O_{12}$ | --- | Temperature: 1500 °C. Flow velocity: 4.4 cm/s. Composition: 50%$H_2O$/10%$O_2$. Testing up to 100 h | -0.2 & -0.4 mg/cm$^2$ after 100 h | -0.002 mg/cm$^2$·h (second measurement not linear) | [16,42] |
| | Oxidation bonded by reaction sintering $Si_3N_4$ substrate at 1500 °C for 2 h in Ar atmosphere | --- | --- | --- | Temperature: 1500 °C. Flow rate: 175 ml/min. Flow velocity: 0.046 cm/s. Composition: 30%$H_2O$/70%$O_2$. Testing up to 50 h | --- | 0.004688 mg/cm$^2$·h | [46] |
| | Cold pressing + sintering at 1600 °C for 12 h in air | --- | --- | --- | Temperature: 1500 °C. Flow rate: 175 ml/min. Flow velocity: 0.046 cm/s. Composition: 30%$H_2O$/70%$O_2$. Testing up to 50-100 h | --- | -0.75 mg/cm$^2$·h | [45] |
| | Cold pressing + sintering at 1400 - 1600 °C | > 99 wt.% YbDS | --- | < 5% | Temperature: 1500 °C. Flow rate: ~290 l/h. Flow velocity: 13 cm/s. Composition 30%$H_2O$/70% air. $P_{H_2O}$ = 0.3 bar, $P_{total}$ = 0.1 MPa. Testing up to 310 h | -0.616 mg/cm$^2$ after 310 h | Not linear | [38] |
| | Si bond coat and YbDS top coat deposited using air plasma spraying on top of SiC substrates | YbDS: 62 wt.% / YbMS: 38% | YbDS: 32 wt.% / YbMS: 68 wt.% | 2 % | Temperature: 1200 °C. Flow velocity: 100 m/s. $P_{H_2O}$ = 0.15 atm, $P_{total}$ = 1 atm. Testing up to 200 h | -0.1 μm/h | --- | [57] |
| | Air plasma spraying + heat treatment in air at 1500 °C for 40 h | YbDS: 70 wt.% / YbMS: 30 wt.% | YbDS: 5 wt.% / YbMS: 95 wt.% | 7 % | Temperature: 1400 °C. Flow velocity: 90 m/s. $P_{H_2O}$ = 0.15 atm, $P_{total}$ = 1 atm. Testing up to 200 h | -0.3 mg/cm2 after 200 h | Not linear | [58] |
| | Spark plasma sintering at 1650 °C/50 MPa in vacuum | YbDS: 92 wt.% / YbMS: 8 wt.% | YbDS: 14 wt.% / YbMS: 86 wt.% | < 2% | | -0.1 mg/cm2 after 200 h | 0.0005 mg/cm$^2$·h | [58] |

| Material | Deposition method | As deposited phase composition | Post-testing phase composition | Porosity | Testing conditions | Max volatilisation | Volatilisation rate | Reference |
|---|---|---|---|---|---|---|---|---|
| Lu$_2$SiO$_5$ (LuMS) | Uniaxial cold pressing (50 MPa) + sintering at 1580 °C for 3 h | LuMS: 88 wt.% / LuDS: 12 wt.% | --- | 1 % | Temperature: 1350 °C. Flow rate: 40 ml/min. Composition: 90%H$_2$O/10%O$_2$. Testing up to 166 h | -0.859 mg/cm$^2$ after 166 h | -0.00596 mg/cm$^2$·h | [48] |
| | Hot pressing at 1500 °C/27.6 MPa in vacuum | LuMS | LuMS. Traces of Al$_5$Lu$_3$O$_{12}$ | --- | Temperature: 1500 °C. Flow velocity: 4.4 cm/s. Composition: 50%H$_2$O/10%O$_2$. Testing up to 100 h | 0.3 & 0.65 mg/cm$^2$ after 100 h | 0.003 & 0.0065 mg/cm$^2$·h | [16] |
| | Dip coating CMC substrate + heat treatment at 1350 °C for 50 h | > 90 wt.% LuMS, < 10 wt.% LuDS, Lu$_2$O$_3$ | Mainly LuDS | ~10% | Temperature: 1350 °C. Flow rate: 0.67 cm$^3$/s. Composition: 90%H$_2$O/10%O$_2$. Testing up to 150 h | 0.69 mg/cm$^2$ after 150 h | 0.00256 mg/cm$^2$·h | [49] |
| Lu$_2$Si$_2$O$_7$ (LuDS) | Oxidation bonded by reaction sintering Si$_3$N$_4$ substrate at 1500 °C for 2 h in Ar atmosphere | LuDS | --- | --- | Temperature: 1500 °C. Flow rate: 175 ml/min. Flow velocity: 0.046 cm/s. Composition: 30%H$_2$O/70%O$_2$. Testing up to 50 h | --- | 0.002218 mg/cm$^2$·h | [46] |
| | Cold pressing + sintering at 1600 °C for 12 h in air | --- | --- | --- | Temperature: 1500 °C. Flow rate: 175 ml/min. Flow velocity: 0.046 cm/s. Composition: 30%H$_2$O/70%O$_2$. Testing up to 50-100 h | --- | -0.0042 mg/cm$^2$·h | [45] |
| | Cold pressing + sintering at 1400 - 1600 °C | > 99 wt.% LuDS | --- | < 5 % | Temperature: 1500 °C. Flow rate: ~290 l/h. Flow velocity: 13 cm/s. Composition 30%H$_2$O/70% air. PH$_2$O = 0.3 bar, Ptotal = 0.1 MPa. Testing up to 310 h | -0.156 mg/cm$^2$ after 310 h | -0.00009 mg/cm$^2$·h | [38] |
| | Hot pressing at 1600 °C/20 MPa for 3 h in Ar atmosphere | LuDS and LuMS | LuDS, LuMS and Lu$_2$O$_3$ | --- | Temperature: 1300 °C. Flow rate: 175 ml/min. Composition: 30%H$_2$O/70% air. Testing up to 100 h | --- | -0.001427 mg/cm$^2$·h | [44] |
| Sc$_2$Si$_2$O$_7$ (ScDS) | Hot pressing at 1500 °C/27.6 MPa in vacuum | ScDS. Traces of SiO$_2$ | ScDS | --- | Temperature: 1500 °C. Flow velocity: 4.4 cm/s. Composition: 50%H$_2$O/10%O$_2$. Testing up to 100 h | -0.4 & -0.45 mg/cm$^2$ after 100 h | Not linear | [16] |

*Table 2: Summary of the volatilisation of different rare earth silicates. The maximum volatilisation was approximated from plots where no explicit data was available. Volatilisation rate was calculated where no explicit rate was provided, assuming linear behaviour.*

The fact that different test conditions were used in most of the experiments summarised in Table 2 makes it difficult to draw conclusions directly from the volatilisation rates. It should be kept in mind that this rate is dependent of the temperature, steam velocity, steam partial pressure and total pressure, as indicated by equations 4 and 5. In order to provide a more comparable quantity, which could be used to assess the resistance to steam volatilisation of different compositions, several approaches have been taken. First, as it was previously mentioned, considerable effort has been placed into developing a theoretical model that can predict this effect [40–42]. This line of work has provided some interesting results, for instance, showing confirmation that as a general characteristic, rare earth monosilicates tend to experience lower volatility rates than their disilicate counterparts. Nevertheless, the current state of the research does not provide a detailed description of the volatility rates to be expected for different compositions at different test conditions, which complicates the comparison. Another approach has been proposed recently, based on more fundamental chemical concepts. Optical basicity (OB or Λ) was first introduced by Duffy and Ingram [67] aiming to classify the chemical activity of oxides in glass, being defined as the ability of oxygen anions to donate electrons, which depends on the polarizability of the metal cations [68]. This chemical criterion has been suggested as a potential quantity useful for comparison between different compositions, as higher optical basicity values correlate to lower steam-induced volatility [1]. This correlation has not yet been confirmed, making comparison of experimental data, such as the one presented in Table 2, still a valuable insight into the volatility of different rare earth silicates.

This volatilisation not only removes material from the top coat, reducing the time required for oxidisers to diffuse to the silicon bond coat, but also can cause the appearance of connected porosity, which represents a preferential pathway for the ingress of oxidisers. This phenomenon, reported by Richards *et al.* [69] on a APS deposited YbDS top coat with a Si bond layer, tested under steam cycling conditions (total pressure of 1 atm, oxygen partial pressure of 0.1 atm, flow velocity of 4.4 cm/s, 90% $H_2O$/10% $O_2$ environment, 60 min at 1316 °C and 10 min at 110 °C for up to 2000 h), is shown below in Figure 11.

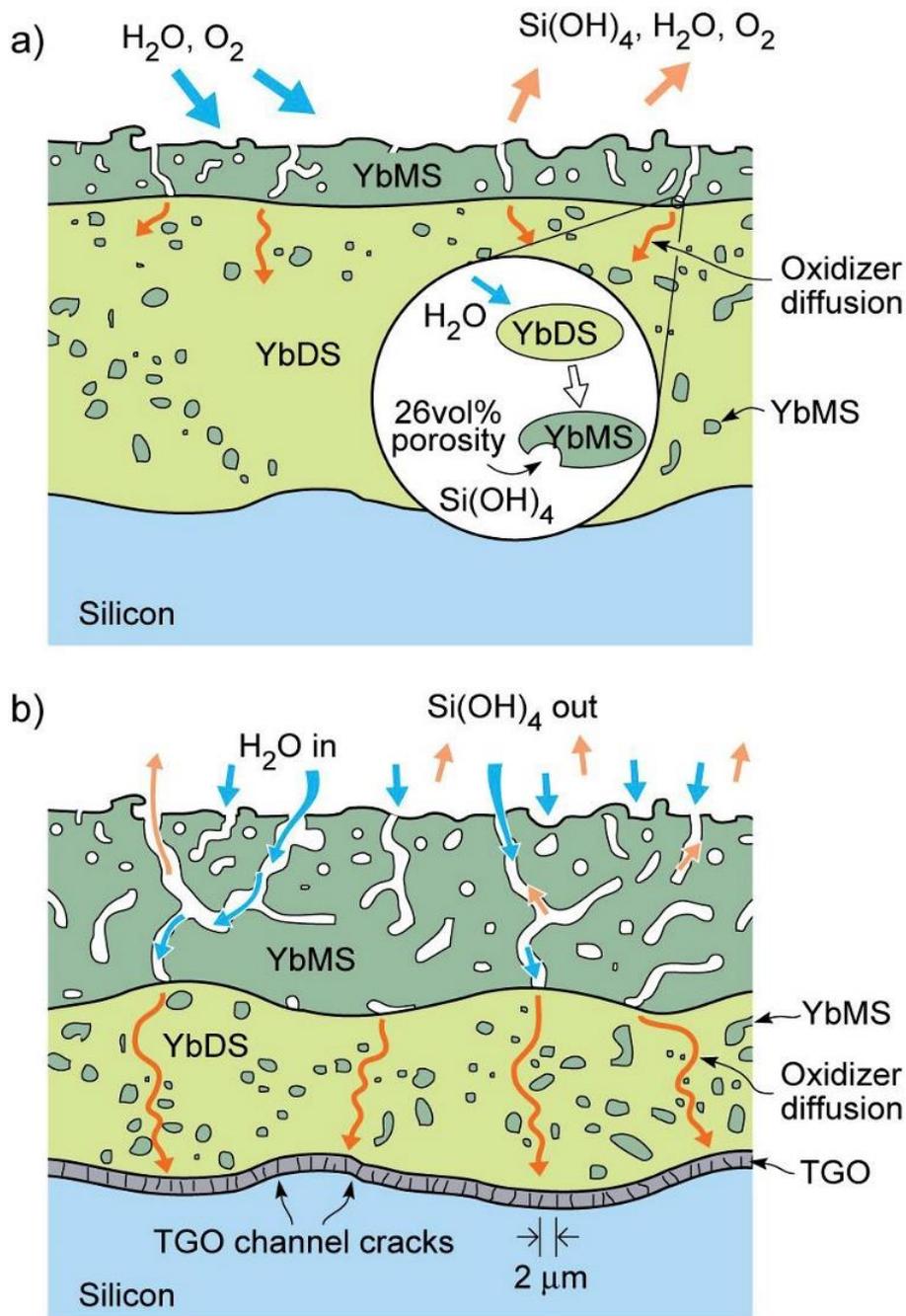

Figure 11: Schematic representation of the volatilisation of silica from the initial YbDS and formation of YbMS. (a) shows the initial stage of the process, while (b) shows the late stages [69]

Assuming that the only volatile product produced is Si(OH)$_4$, the transformation from YbDS into YbMS, described in Equation 7, implies a volume reduction of 26%. This coupled with the increased content of YbMS (which has a higher CTE when compared to SiC) produced a CTE mismatch that induced thermal stresses upon cycling testing, producing vertical cracks and facilitating the access of oxidisers to the silicon bond coat. This preferential access of oxidisers to the silicon bond coat produced a quick growth of the TGO.

$$2Yb_2Si_2O_7(s) + 2H_2O(g) \rightarrow Yb_2SiO_5(s) + Si(OH)_4(g) \tag{7}$$

On the other hand, the formation of a monosilicate layer on top of the disilicate can act as passivation barrier, due to the lower volatilisation of monosilicates when compared to disilicates, associated with a lower silica activity, as discussed in the section "Second generation". Although this passivation layer can reduced the volatility rate of the coating, excessive formation of porosity and high steam flow velocities can cause the erosion of these layers, effectively increasing the mass loss rate [57].

In addition to the volatilisation of the rare earth silicate, failure of EBCs exposed to steam containing environments can take place due to the appearance of vertical cracks and spallation. Regarding vertical cracks, they can be formed due to CTE mismatch between the initial compositions of the different layers, as mentioned before, or due to the formation of a new phase with a different CTE value. This situation may arise in the case of top layers made of rare earth silicates with several polymorphs. The newly formed cracks allow the access of oxidisers to the silicon bond coat, inducing the rapid growth of a β-cristobalite TGO. Upon cooling below ~220 °C, this β-cristobalite $SiO_2$, transforms to the α-phase, process accompanied by a volume reduction of approximately 4.5% [55,56]. This process promotes the formation of cracks parallel to the interface, which eventually lead to the coating spallation, as shown in Figure 12a. Figure 12b shows the mentioned change of the CTE of cristobalite with temperature, seeing a sharp change around ~220 °C (~500 K).

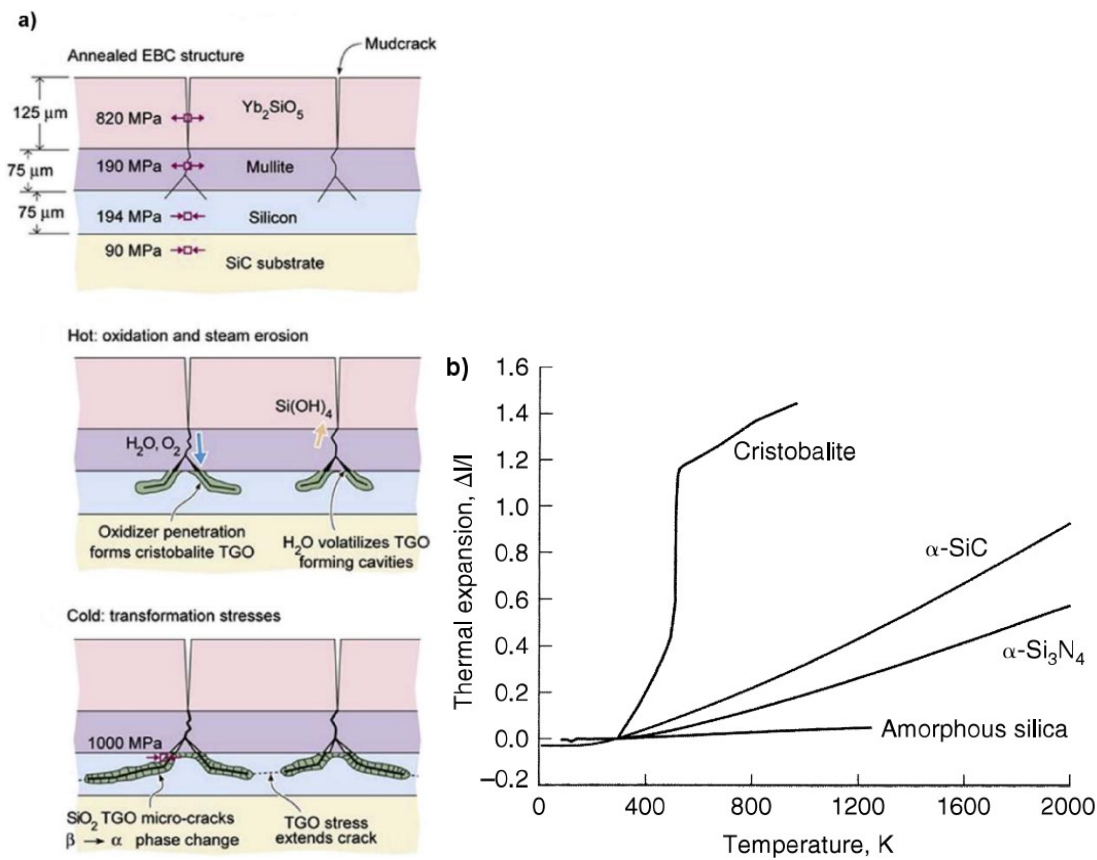

*Figure 12: a) Schematic of the spallation process induced on multilayer EBC systems due to the formation of β-cristobalite TGO and the transformation to α-phase upon cooling, causing cracking. b) Change of the coefficient of thermal expansion with temperature on cristobalite, SiC, $Si_3N_4$ and amorphous silica [56] [9]*

The results presented in this section support the idea that the design and study of the performance of EBCs should be approached as a multifaceted problem. A low volatilisation rate is not enough for a composition to be considered as the optimal EBC top coat, since CTE matching, chemical stability and phase transformation at high temperature also play an essential role. Results coming from approximate models, such as sintered bodies, will still provide useful knowledge, but if a successful transition to real world applications is to be achieved, further testing with production-like deposition methods and testing is required to understand the fundamental mechanisms of steam corrosion on EBC. Even when realistic coatings are produced and tested under the appropriate conditions, attention should be paid not only to one single phenomenon involved in the failure of the coating. That is not to say that single phenomenon should not be thoroughly investigated, as a deeper understanding of the causes will allow for better performing coatings, but it should be kept in mind that a compromise between the requirements is needed for proper performance during service. Top coat volatilisation is, undoubtedly, a serious issue, but it is only one of the potential failure modes. Cracking due to CTE mismatch and oxidation of the Si bond coat, leading to detrimental phase transformation within the TGO are also important occurrences that need to be considered when designing the test methodology. Finally, although the isolation of the effect that steam has on EBCs is needed to understand the basis of its attack, it should not be forgotten that steam is not the only component present in the environment experienced by EBCs during service. Molten salts, or CMAS, as it is discussed in the section "CMAS corrosion" represent a severe challenge for the current iteration of EBCs, and steam protection alone will not suffice for the successful application of rare earth silicates.

### 3.2. CMAS corrosion

During the early development of first generation of EBCs, the main concern was the degradation suffered by the SiC CMC substrates by molten alkali salts. Due to the variability in specific compositions, the term CMAS (CaO-MgO-$Al_2O_3$-$SiO_2$) will be used to denote the multitude of impurities that represent a threat when ingested by the engine or turbine. As previously mentioned in this work, that focus shifted towards steam once it was realised that steam presented also a considerable threat to the performance and service life of SiC components. Nevertheless, CMAS was, and still is, a crucial obstacle that any potential EBC system must surpass, and research has continued in this regard trying to understand the interaction between molten CMAS and EBCs. In this work only research done on rare earth silicates will be presented.

One of the particularities of the interaction of CMAS with coatings is that temperature plays a critical role. Not by accident, in this work the description of salts or CMAS has been always accompanied by "molten". In its many configurations, CMAS does not represent a problem as long as it remains in solid form. Although the exact melting point for CMAS varies with the precise composition used, the commonly agreed melting point for CMAS is ~1200 °C, well below the service temperature at which EBCs are expected to operate, of ~1500 °C. The problem is not new, as CMAS has been a thoroughly investigated topic in relation to YSZ coatings for thermal barrier coating (TBC) applications [70–74]. The corrosion pathways are, however, different in the case of rare earth silicates, requiring of additional research. This provides an additional challenge, as the CMAS composition is highly variable, as mentioned previously, and different compositions have been demonstrated to present different reactions [75,76], as shown in Figure 13.

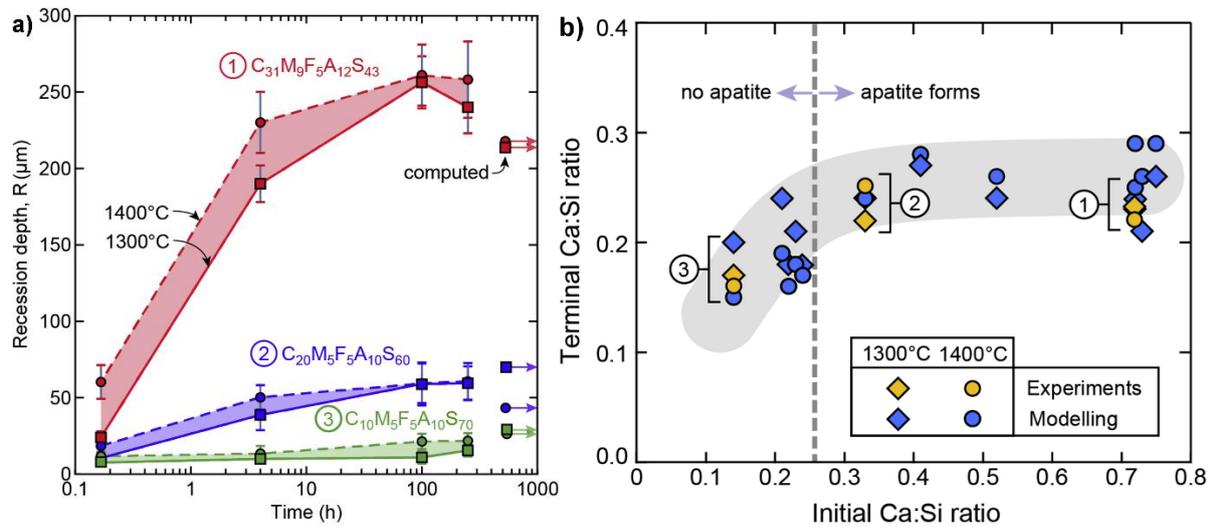

*Figure 13: a) Experimental data on recession depth of the YDS surface after CMAS attack with different compositions. Theoretically derived values are represented to the right as "computed" (1) indicates a CaO rich composition, (2) an intermediate CaO composition and (3) a CaO lean composition. b) Terminal Ca:Si ratio versus initial Ca:Si after heat treatment of YDS for two temperatures and three CMAS compositions, both from theoretical calculations and experimental results. The dashed grey line represents the minimum initial Ca:Si ratio discovered for the formation of apatite precipitates* [77]

Another factor that should be considered is that the majority of the studies regarding rare earth silicates and CMAS have been reported on sintered pellets or bulk material. Sintered bodies or bulk materials, as discussed with the steam interaction, can provide useful information, but should be treated carefully if conclusions are to be extracted regarding coatings produced through thermal spraying, as required per many sectors of the industry. Additionally, these studies tend to be performed in phase-pure fully crystalline materials, which does not accurately represent the reality of deposited coatings. The differences between tests performed on sprayed coatings and sintered bodies are clearly represented in Figure 14.

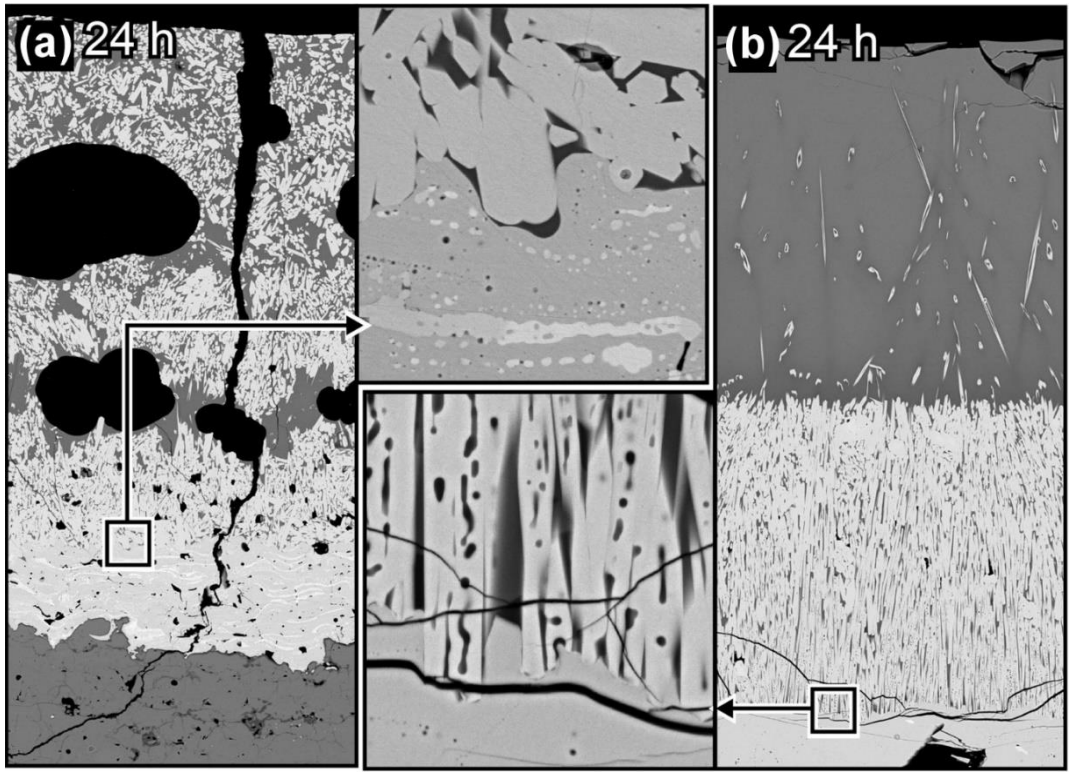

*Figure 14: YDS and CMAS interaction after 24 h at 1300 °C in air where an APS-deposited coating (a) is shown versus a sintered body (b). The same CMAS composition and testing conditions were used in both cases. Modified from [78]*

Although the degradation suffered by EBCs due to CMAS is highly variable depending on the testing temperature, the composition of the CMAS used, the deposition method chosen and the composition of the EBC top coat (pure phase or mixed phases), as mentioned above, several common aspects have been discovered when studying the interaction of rare earth silicates and CMAS at high temperatures. Two main degradation mechanisms have been identified, with examples being shown in Figure 15. The first interaction observed involves the reaction between the molten CMAS and the EBC top coat. Such mechanism has been reported for $Y_2SiO_5$ and $Y_2Si_2O_7$ [15,75,77,79–83], in which the reaction with the CMAS produces the dissolution of the EBC followed by the recrystallisation of yttrium monosilicate and Y-Ca-Si apatite in solid solution, forming characteristic needle-like structures, as it can be seen in Figure 15a. The second possibility is based not on the reaction between CMAS and the top coat, but on the penetration of the CMAS material along grain boundaries, reaching deeper layers of the EBC and causing "blister" damage, as seen in Figure 15b for $Yb_2Si_2O_7$, due to the dilatation gradient caused by the slow penetration of CMAS.

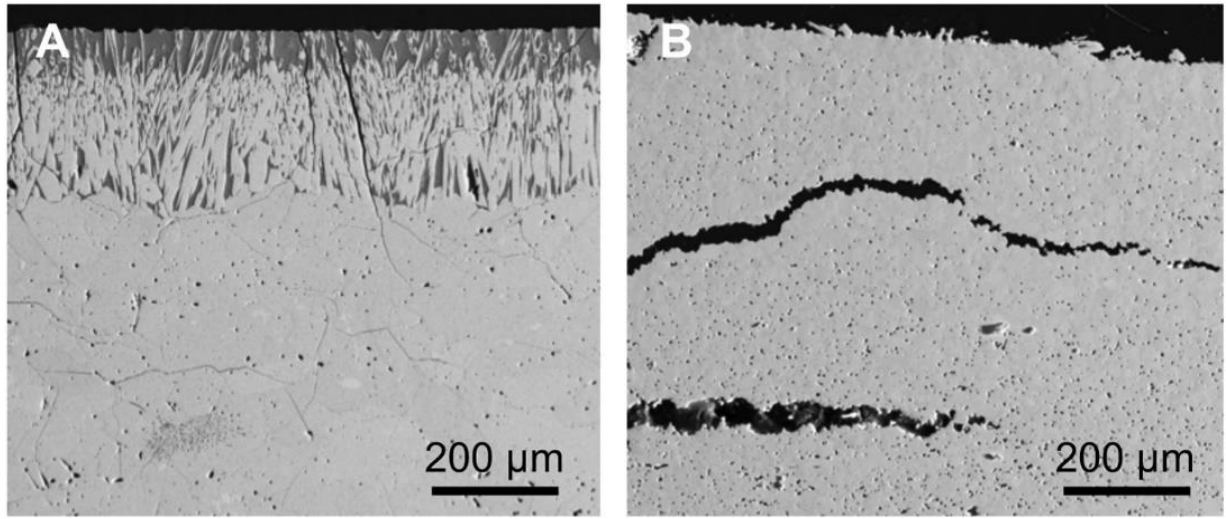

*Figure 15: Cross-section SEM images of rare earth silicates pellets exposed to CMAS at 1500 °C for 24 h. Image A corresponds to $Y_2Si_2O_7$ and image B to $Yb_2Si_2O_7$. Adapted from* [36]

In addition to the chemical composition of the EBC top coat, the morphology of the coating (due to the deposition method chosen) and the presence of a mixture of phases also affects the degradation mechanism present. For instance, for $Yb_2SiO_5$ and $Yb_2Si_2O_7$ [75,79,80,84–89], there seems to be a difference in the mechanism involved when the EBC is exposed to CMAS attack at high temperatures depending on whether the testing involves sintered bodies or thermal sprayed coatings. Some authors report minimal reaction between the Yb mono- and di-silicate pellets, rather showing intensive penetration of CMAS along grain boundaries. Nevertheless, several studies on thermal sprayed coatings have shown extensive dissolution of the ytterbium silicate and the precipitation of needle-like apatite structures, much like with yttrium silicates, as it can be seen in the schematic proposed by Zhao *et al.* [86] for the mechanism taking place, shown in Figure 16. Both this work, and the ones conducted by Stolzenburg *et al.* [87] and Poerschke *et al.* [78] deserve special attention as they were performed on APS deposited coatings, which provides a unique perspective not fully captured with sintered pellets studies.

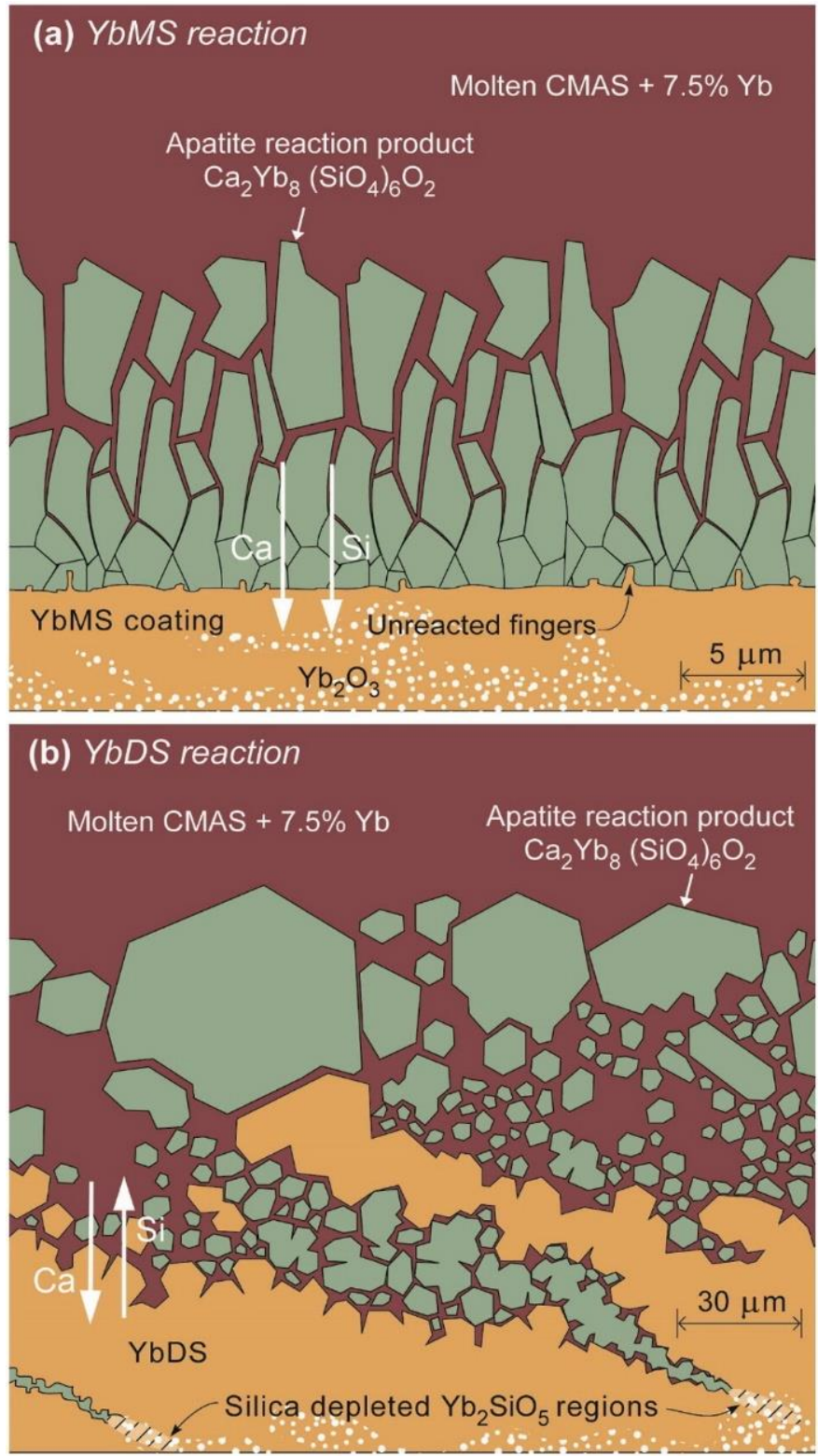

*Figure 16: Schematic of the proposed interaction mechanism between YbMS (top) and YbDS (bottom) when exposed to CMAS at 1300 °C* [86]

The mechanism shown in Figure 16a) for YbMS is based in the discussed dissolution of the monosilicate and posterior precipitation as needle-like apatite grains with areas of intercalated residual CMAS, forming the already seen reaction layer. In the case of YbDS, shown in Figure 16b), the initial apatite grains are coarser and larger in size, with an irregular reaction layer where no clear reaction front can be determined. As the corrosion continues, molten CMAS preferentially attacks the YbMS-rich areas of the coatings. Due to the lamellar structure of the plasma deposited coatings, the YbMS-rich splats are elongated and parallel to the surface, which produces a rapid advance of the reaction in this direction. The precipitation of the apatite grains creates a "cleft" effect, as it can be seen in the schematic.

This effect is one of the most clear examples reported of the different corrosion attacks mechanism that can be shown in deposited coatings versus sintered bodies, since the latter tends to be a pure phase without presence of splats or enriched and leaner areas. This variability depending of the experimental methods applied and the testing conditions is presented in Table 3, where a summary of different CMAS corrosion experiments is presented, attending to the deposition techniques for both the substrate and the CMAS, the specific CMAS composition used and the testing conditions (such as temperature, CMAS mass loading or high temperature exposure time).

| Material | CMAS composition (mol %) | CMAS preparation | Test material preparation | Testing conditions | Corrosion effects | Ref. |
|---|---|---|---|---|---|---|
| Yb$_2$SiO$_5$ (YbMS) | 35CaO-10MgO-7AlO$_{1.5}$-48SiO$_2$ | CMAS + ethanol applied as paste | Hot pressing at 1500 °C/103 MPa in vacuum | ~40 mg/cm$^2$ 1500 °C for 50h | Preferential attack at grain boundaries. 4 mm CMAS penetration. Reaction layer with hexagonal-shaped apatite grains | [80] |
| | 33CaO-9MgO-13AlO$_{1.5}$-45SiO$_2$ | 100 °C for 10 h + 1200 °C for 24 h + cold pressing + sintering at 1200 °C for 2 h | Sol-gel + cold pressing + sintering at 1500 °C for 10 h in air | 1200 °C for 4 h 50%H$_2$O/50%O$_2$ | Reaction layer at the CMAS/YbMS interface, with the presence of Yb-doped CaAl$_2$Si$_2$O$_8$ | [90] |
| | | 1200 °C for 20 h + cold pressing | Air plasma sprayed YbMS/mullite/Si on SiC substrates + 1300 °C for 20 h | 1300 °C for 250 h | Apatite reaction layer after 1 minute with vertical needle-like grains. Thicker reaction layer and grain coarsening with increasing time. EBC fully penetrated after 250 h, layers reacted forming large pores | [86] |
| | | 1550 °C for 4 h twice. CMAS and YbMS powders mixed 70:30 wt.% | Commercially available | 1300 °C for 96 h | YbMS phase content dropped to 27% after 1 minute, dropping to 7% after 96 h. Apatite appears in its place forming needle-like hexagonal precipitates | [84] |
| | | 1550 °C for 4 h twice. Placed on a well on the bulk YbMS | Not reported | ~35 mg/cm$^2$ 1300 °C for 96 h | Extensive reaction between the bulk YbMS and the molten CMAS to form needle-like hexagonal apatite precipitates dispersed along the residual CMAS | |
| Yb$_2$Si$_2$O$_7$ (YbDS) | 33CaO-9MgO-13AlO$_{1.5}$-45SiO$_2$ | 100 °C for 10 h + 1200 °C for 24 h + cold pressing + 1200 °C for 2 h | Sol-gel + cold pressing + 1500 °C for 10 h | ~314 mg/cm$^2$ 1400 °C for 10 h 50%H$_2$O/50%O$_2$ | Preferential attack at grain boundaries. 1.5 - 2 μm CMAS penetration. Formation of large pores | [79] |
| | 39.2CaO-5.2MgO-4.1AlO$_{1.5}$-51.5SiO$_2$ | 1550 °C for 4 h twice. CMAS + ethanol applied as paste | Spark plasma sintering at 1600 °C/75 MPa + 1500 °C for 1 h | ~15 mg/cm$^2$ 1500 °C for 24 h | Dense CMAS glass layer with apatite grains after 1 h, both hexagonal and needle-like. After 24 h severe blister damage is seen, almost no presence of apatite. CMAS penetrating through grain boundaries | [81] |
| | 30.7CaO-8.2MgO-12.8AlO$_{1.5}$-48.3SiO$_2$ 24.8CaO-9.1MgO-14.2AlO$_{1.5}$-51.7SiO$_2$ 6.7CaO-8.9MgO-14.2AlO$_{1.5}$-70.1SiO$_2$ | 850 °C for 10 h + 1500 °C for 1 h + water quenching. 50:50 mol% EBC:CMAS | Commercially available + 1500 °C for 10 h | 1200 °C, 1300 °C and 1400 °C for 1 h | At the highest CaO content there was formation of apatite at all three temperatures, YbDS still present. At 1200 °C there is little presence of needle-like apatite | [75] |
| | 33CaO-9MgO-13AlO$_{1.5}$-45SiO$_2$ | 1200 °C for 20 h + cold pressing | Air plasma sprayed YbDS/mullite/Si on SiC substrates + 1300 °C for 20 h | 1300 °C for 250 h | For lower CaO contents, YbDS was produced at all temperatures along with cristobalite | [86] |
| | | 1550 °C for 4 h twice. CMAS and YbMS powders mixed 70:30 wt.% | Commercially available | 1300 °C for 96 h | Irregular apatite reaction layer after 1 h with coarse grains. No clear reaction front seen, CMAS preferentially reacted with YbMS-rich areas, creating quickly advancing fronts acting as clefts | [84] |
| | | 1550 °C for 4 h twice. Placed on a well on the bulk YbDS | Not reported | ~35 mg/cm$^2$ 1300 °C for 96 h | YbDS phase content dropped to 30% after 96 h. No apatite is detected after 96 h, some dissolution of the YbDS | |

| Material | CMAS composition (mol %) | CMAS preparation | Test material preparation | Testing conditions | Corrosion effects | Ref. |
|---|---|---|---|---|---|---|
| $Y_2SiO_5$ (YMS) | $33CaO-9MgO-13AlO_{1.5}-45SiO_2$ | 1200 °C for 24 h + cold pressing + 1220 °C for 2 h | Pellets provided by industrial partner | ~13 mg/cm$^2$ 1300 °C for 100 h | 80 µm of recession after 100 h. Needle-like apatite grains reaction layer | [15] |
| | $33CaO-9MgO-13AlO_{1.5}-45SiO_2$ | 100 °C for 10 h + 1200 °C for 24 h + cold pressing + 1200 °C for 2 h | Sol-gel + cold pressing + 1500 °C + 10 h | 1200 °C for 4 h 50%H$_2$O/50%O$_2$ | Reaction layer at the CMAS/YMS interface, with the presence of Y-doped $CaAl_2Si_2O_8$ | [90] |
| $Y_2Si_2O_7$ (YDS) | $35CaO-10MgO-7AlO_{1.5}-48SiO_2$ | CMAS + ethanol applied as paste | Hot pressing 1500 °C/103 MPa | ~40 mg/cm$^2$ 1500 °C for 50h | Preferential attack at grain boundaries. 4 mm CMAS penetration. Reaction layer with apatite needle-like grains and Si-rich areas | [80] |
| | $33CaO-9MgO-13AlO_{1.5}-45SiO_2$ | 100 °C for 10 h + 1200 °C for 24 h + cold pressing + 1200 °C for 2 h | Sol-gel + cold pressing + 1500 °C + 10 h | ~314 mg/cm$^2$ 1400 °C for 10 h 50%H$_2$O/50%O$_2$ | Dense apatite reaction layer with minimal CMAS penetration | [79] |
| | $39.2CaO-5.2MgO-4.1AlO_{1.5}-51.5SiO_2$ | 1550 °C for 4 h twice. CMAS + ethanol applied as paste | 1600 °C for 4 h + spark plasma sintering at 1600 °C/75 MPa + 1500 °C for 1 h | ~15 mg/cm$^2$ 1500 °C for 24 h | 300 µm apatite reaction zone after 24 h with 2 layers: (1) needle-like grains and CMAS and (2) dense apatite grains | [81] |
| | $25.2CaO-2.6MgO-8.2AlO_{1.5}-59.8SiO_2-1.6FeO_{1.5}-1.5K_2O$ | Placed on a well on the bulk YDS | Commercially available + hot pressing 1500 °C/27.6 MPa for 2 h | ~35 mg/cm$^2$ 1200 °C, 1300 °C, 1400 °C and 1500 °C for 20 h | ~215 µm of penetration at 1500 °C for 20 h, reaction zone with 2 layers: (1) apatite grains and CMAS (2) needle-like apatite grains with new pores. Grains in (1) transition to needle-like with increasing temperature | [91] |
| | $30.7CaO-8.2MgO-12.8AlO_{1.5}-48.3SiO_2$ $24.8CaO-9.1MgO-14.2AlO_{1.5}-51.7SiO_2$ $6.7CaO-8.9MgO-14.2AlO_{1.5}-70.1SiO_2$ | 850 °C for 10 h + 1500 °C for 1 h + water quenching. 50:50 mol% EBC:CMAS | Commercially available + 1500 °C for 10 h | 1200 °C, 1300 °C and 1400 °C for 1 h | For the highest CaO content, formation of apatite, grain size increases with temperature. For reduced CaO content, unreacted YDS and cristobalite are detected. For the lowest CaO content, no apatite detected, only crystallised CMAS | [75] |
| | $31CaO-9MgO-5FeO_{1.5}-12AlO_{1.5}-43SiO_2$ | ~50 °C below melting point for 24 h + cold pressing + 1100 °C for 12 h | Powder provided by industrial partner + field-assisted sintering at ~1500 °C/~100 MPa + 1400 °C for 24 h | ~15 mg/cm$^2$ 1300 °C for 24 h | Recession of 25 µm with ~15 µm reaction layer with needle-like apatite grains after 10 min. After 4 h recession is 180 µm with thicker reaction layer and cristobalite. After 24 h recession is 220 µm with thicker reaction layer due to growth and formation of needle-like apatite grains | [78] |
| | | | Air plasma sprayed YDS /Si on CMC substrates provided by industrial partner + 1325 °C for 10 h | ~15 mg/cm$^2$ 1300 °C for 100 h | Recession of 60 µm with coarse apatite grains after 10 min. After 4 h recession is 150 µm with needle-like grains, CMAS with pores and dispersed apatite grains. After 24 h recession is 200 µm with coarser grains and more pores and grains in the CMAS. Cracks reaching substrate appear. After 100 h recession is ~250 µm | |
| | $31CaO-9MgO-5FeO_{1.5}-12AlO_{1.5}-43SiO_2$ $20CaO-5MgO-5FeO_{1.5}-10AlO_{1.5}-60SiO_2$ $10CaO-5MgO-5FeO_{1.5}-10AlO_{1.5}-70SiO_2$ | ~50 °C below melting point for 24 h + cold pressing + 1100 °C for 12 h | Field-assisted sintering at 1470 °C/100 MPa + 1400 °C for 24 h | ~18 mg/cm$^2$ 1300 °C and 1400 °C for 250 h | For intermediate and lowest CaO content CMAS shows large pores after 10 min. Recession was maximum after 100 h at 1300 °C, being ~248, 59 and 16 µm from higher to lower CaO content. Slightly faster recession but similar final values for 1400 °C | [77] |

| Material | CMAS composition (mol %) | CMAS preparation | Test material preparation | Testing conditions | Corrosion effects | Ref. |
|---|---|---|---|---|---|---|
| $Lu_2SiO_5$ (LuMS) | 33CaO-9MgO-13AlO$_{1.5}$-45SiO$_2$ | 100 °C for 10 h + 1200 °C for 24 h + cold pressing + 1200 °C for 2 h | Sol-gel + cold pressing + 1500 °C for 10 h | 1200 °C for 4 h 50%H$_2$O/50%O$_2$ | Reaction layer at the LuMS/CMAS interface, presence of Lu-doped CaAl$_2$Si$_2$O$_8$ | [90] |
| $Lu_2Si_2O_5$ (LuDS) | | | | ~314 mg/cm$^2$ 1400 °C for 10 h 50%H$_2$O/50%O$_2$ | Preferential attack at grain boundaries, 2 μm CMAS penetration after 10 h. Formation of large pores | [79] |
| $La_2SiO_5$ (LaMS) | | | | 1200 °C for 4 h 50%H$_2$O/50%O$_2$ | Reaction layer at the LaMS/CMAS interface with dendritic, tree-like precipitates. La$^{3+}$ cations diffuse easily into CMAS | [90] |
| $La_2Si_2O_5$ (LaDS) | | | | ~314 mg/cm$^2$ 1400 °C for 10 h 50%H$_2$O/50%O$_2$ | Formation of branch shaped, tree-like crystals. Presence of Ca$_3$La$_6$(SiO$_4$)$_6$ and Ca$_3$La$_8$(SiO$_4$)$_6$O$_2$ | [79] |
| $Gd_2SiO_5$ (GdMS) | | | | 1200 °C for 4 h 50%H$_2$O/50%O$_2$ | Reaction layer at the GdMS/CMAS interface with presence of dendritic, tree-like precipitates | [90] |
| $Gd_2Si_2O_7$ (GdDS) | | | | ~314 mg/cm$^2$ 1400 °C for 10 h 50%H$_2$O/50%O$_2$ | Discontinuous reaction layer of apatite | [79] |
| | 30.7CaO-8.2MgO-12.8AlO$_{1.5}$-48.3SiO$_2$ 24.8CaO-9.1MgO-14.2AlO$_{1.5}$-51.7SiO$_2$ 6.7CaO-8.9MgO-14.2AlO$_{1.5}$-70.1SiO$_2$ | 850 °C for 10 h + 1500 °C for 1 h + water quenching. 50:50 mol% EBC:CMAS | Commercially available + cold pressing + 1580 °C for 10 h | 1400 °C for 1 h | At the lowest CaO content there is apatite precipitates along with cristobalite. Highest CaO content produces needle-like apatite precipitates | [76] |
| $Eu_2SiO_5$ (EuMS) | 33CaO-9MgO-13AlO$_{1.5}$-45SiO$_2$ | 100 °C for 10 h + 1200 °C for 24 h + cold pressing + 1200 °C for 2 h | Sol-gel + cold pressing + 1500 °C for 10 h | 1200 °C for 4 h 50%H$_2$O/50%O$_2$ | Reaction layer at the EuMS/CMAS interface with dendritic, tree-like precipitates | [90] |
| $Eu_2Si_2O_7$ (EuDS) | | | | ~314 mg/cm$^2$ 1400 °C for 10 h 50%H$_2$O/50%O$_2$ | Dense apatite reaction layer with the presence of clefts or blister damage that could lead to spallation | [79] |
| $Sc_2Si_2O_7$ (ScDS) | | | | ~314 mg/cm$^2$ 1400 °C for 10 h 50%H$_2$O/50%O$_2$ | Thin, dense apatite reaction layer with 500 μm of CMAS penetration | [79] |
| | 39.2CaO-5.2MgO-4.1AlO$_{1.5}$-51.5SiO$_2$ | 1550 °C for 4 h twice. CMAS + ethanol applied as paste | 1600 °C for 4 h + spark plasma sintering at 1600 °C/75 MPa + 1500 °C for 1 h | ~15 mg/cm$^2$ 1500 °C for 24 h | Dense residual CMAS glass with scattered apatite grains after 1 h. After 24 h sever blister damage with no presence of apatite. Reaction between ScDS and CMAS, with Sc-doped CMAS penetrating through grain boundaries | [81] |
| $Dy_2Si_2O_7$ (DyDS) $Er_2Si_2O_7$ (ErDS) $Nd_2Si_2O_7$ (NdDS) | 30.7CaO-8.2MgO-12.8AlO$_{1.5}$-48.3SiO$_2$ 24.8CaO-9.1MgO-14.2AlO$_{1.5}$-51.7SiO$_2$ 6.7CaO-8.9MgO-14.2AlO$_{1.5}$-70.1SiO$_2$ | 850 °C for 10 h + 1500 °C for 1 h + water quenching. 50:50 mol% EBC:CMAS | Commercially available + cold pressing + 1580 °C for 10 h | 1400 °C for 1 h | Formation of apatite with different stoichiometries and cristobalite presence for all CaO contents. Highest CaO content produces needle-like apatite grains | [76] |

*Table 3: Summary of the CMAS corrosion experimental results reported in the literature for different rare earth silicates*

Despite the wide range of effects described in Table 3, accounting for the variability in rare earth silicates tested, the different CMAS compositions and testing conditions, some general features can be observed. First of all, if reaction between the rare earth silicate and the CMAS does occur, precipitation of RE-Ca-Si apatite is the most common product, with the appearance in occasions of $\beta$-$SiO_2$ cristobalite. This reaction will produce the recession of the coating and appearance of defects such as cracks or porosity. However, reaction with CMAS is not always guaranteed, and penetration of CMAS can also take place without almost interaction, particularly for lower Ca-containing CMAS compositions. This infusion of CMAS into the EBC is undoubtedly undesirable, as it can lead to blister damage as shown in Figure 15b. Secondly, as mentioned before, the CaO content present in the chosen CMAS composition plays a key effect in the corrosion mechanism observed and its severity. Higher CaO contents will have a more nefarious interaction with the top coat in terms of recession rates and precipitation of apatite. On the other hand, lean CaO compositions are still highly undesired due to the potential switch from silicate-CMAS reaction to CMAS penetration, as shown in Figure 13b.

As with the case of steam volatilisation, there is an interest to draw comparison between different rare earth silicate compositions and their experienced CMAS corrosion in order to assess which one might be optimal for the application desired. The use of optical basicity, first introduced in this work for the steam volatilisation, has also been suggested as a rough screening parameter for CMAS resistance [81,88]. The basis behind this criterion is the reduced reactivity between a crystalline oxide ceramic and an oxide glass if their respective OB values are close in value. Although this consideration have value in the initial stages of the EBC design, aiding to choose a composition that in theory could present improved resistance against CMAS corrosion, still presents a rough criterion, which should not be considered to withstand under all conditions and CMAS compositions. Particularly, regarding CMAS compositions, Krause *et al.* [73] reported how the OB values can vary with the specific CMAS compositions, with values ranging between 0.49 to 0.75. This provides another degree of complexity, as debris ingested by engines during service might have different sources, and therefore different interaction with the EBC.

Despite considerable research being conducted regarding the interaction between CMAS and EBC at high temperatures, the fundamental mechanisms that control the interaction are not fully understood yet. One of the reasons, as previously highlighted, is the difference in materials employed and testing protocols. A standardised protocol involving the deposition method used, the acceptable ranges for the phases present in the coating and the deposited microstructure would be needed to fully determine which potential candidate has the best characteristics to provide reliable protection to SiC CMC coated components during operation that involves the ingestion of salt-containing debris. It should be taken into account, however, that a standardised CMAS composition and testing protocol will only be useful for comparison purposes, since the industry might still request specific CMAS compositions, more suited to the debris involved when operating in different areas. A great example of this is the extensive research that was conducted after the eruption in 2010 of the volcano Eyjafjallajökull in Iceland [92–94], representing an unique challenge for aviation in the European air space.

Perhaps, as it was mentioned in the case of the steam corrosion of rare earth silicates, it is too ambitious to expect the same top coat composition to provide effective protection against steam at high flow speeds while showing appropriate CMAS corrosion against a wide range of compositions. To this end, different alternatives are being considered [85,95], and some of them are borrowed from the previous knowledge gathered in the field of TBC, as it is shown in Figure 17.

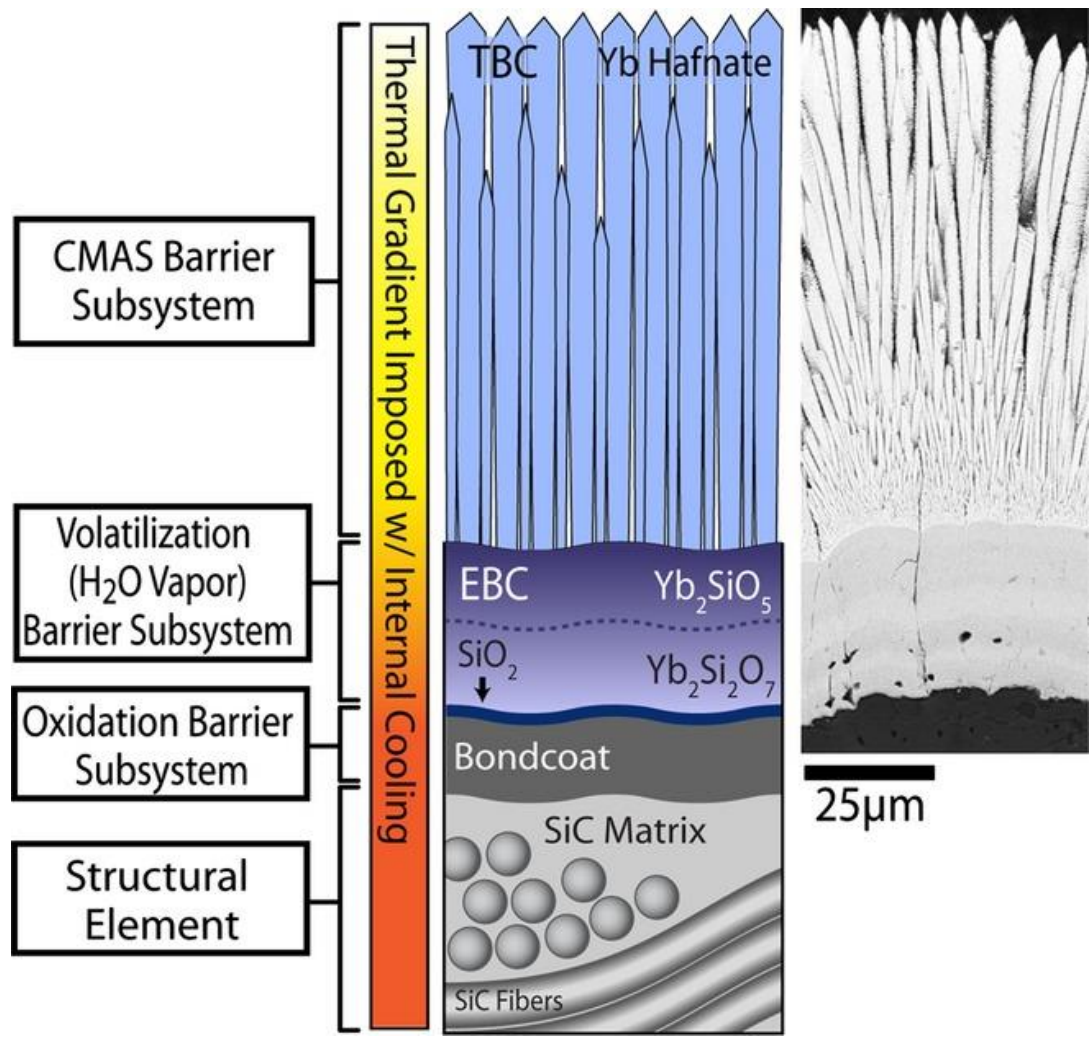

Figure 17: Schematic structure of an EBC sharing features of a thermal barrier coatings (TBC), being mixed TBC/EBC systems aiming to provide protection against steam and CMAS attack [85]

The incorporation of an additional layer to the EBC system might be the way forward to provide complete protection to the component underneath, both from the environment (which could have steam, salt-containing debris or a mixture of both) and from the high temperatures required for a benefit in efficiency. A clear point can be concluded from the current information on CMAS: molten alkali salts represent a formidable challenge that still cannot be tackled with the current generation of EBCs. Recession rates and penetration depths comparable to the standard thickness of EBC systems are observed after a few hundred hours, which is unacceptable if the same situation were to be encountered during service. On top of that, the majority of the reported experiments were conducted with a single application of CMAS, whereas during operation, engines might ingest salt-containing debris continuously, adding fresh molten CMAS to the reaction zone, and preventing the hindering of the nefarious effects due to exhaustion of the components.

# 4. Next generation of EBC

The search for a better performing EBC is never finished, and despite the recent successes that rare earth silicates have collected in terms of protection against the environment and molten salts attacks, there are always new approaches being researched, new pathways towards the next generation of environmental barrier coatings. This section aims to present some of the latest developments in the field of EBC, pointing out potential new avenues that must be further explored before they can be implemented by the industry. In this section different developments are presented, under the common criterion of an EBC that presents a composition beyond the already discussed single rare earth silicate.

As it has been previously said in this work, one of the basic functions of a successful EBC is to provide protection against the environment at which it will operate during service. Regarding the presence of steam and the proven detrimental effect that it has on SiC CMC components, this implies a gas-tight coating capable of preventing the ingress of steam to the substrate underneath. Therefore, it is quite clear that cracks within the coating are highly undesirable, as they represent a preferential pathway for steam to reach the substrate. Nevertheless, cracks are likely to appear during service due to the presence of several temperature cycles, which will cause the accumulation of thermal stresses and, eventually, relaxation vertical cracks. Aiming to increase the service life of EBC systems, research has been carried out with the goal to incorporate self-healing capabilities. The interest for a self-healing EBC can be demonstrated by the presence of patents on the topic [96,97] and the published papers exploring different compositions and mechanisms, as described below. An early example of a self-healing EBC is reported by Nguyen *et al.* [98,99]. A schematic of the proposed mechanism behind the self-healing capabilities can be seen in Figure 18.

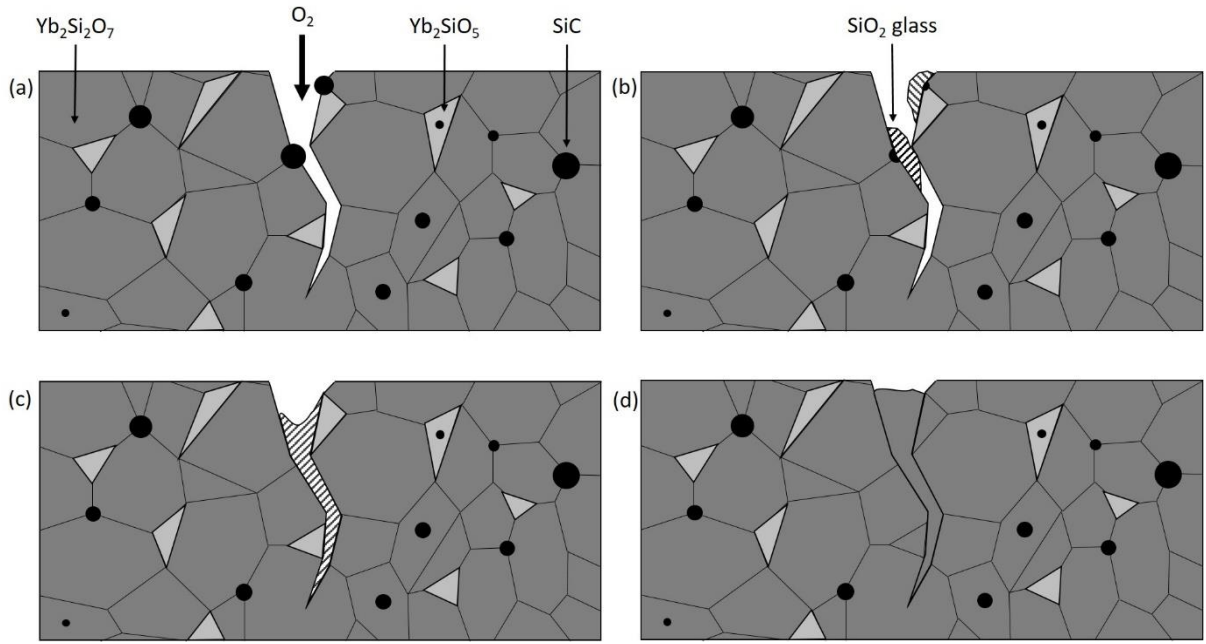

*Figure 18: Schematic of the self-healing mechanism for YbDS/YbMS+SiC systems. Image (a) shows the cracked material, image (b) shows the formation of $SiO_2$ glass (dashed areas) during annealing due to the reaction between the SiC nanoparticles (black circles) and the ingressed oxygen, image (c) shows the filling of the crack with the viscous $SiO_2$ glass and image (d) shows the healing of the crack through the reaction between the $SiO_2$ glass and the YbMS grains (light grey), forming newly created YbDS (dark grey). Redrawn from [99]*

The proposed mechanism is based on the addition of 10 vol.% β-SiC nanocomposites (nanoparticulates, nanofibers or nanowhiskers) to the YbDS/YbMS system. Once a crack appears, as seen in Figure 18a, it provides access for the atmospheric oxygen into the material. During annealing at 1250 °C, the SiC nanofillers react with the atmospheric oxygen, creating viscous $SiO_2$ glass. This viscous amorphous $SiO_2$ is capable of filling the cracks, and then reacts with the YbMS present within the system to form YbDS, effectively sealing the crack due to the associated volume expansion. A similar approach was recently reported by Vu *et al.* [100], where crack self-healing was demonstrated on sintered bodies of composition yttrium monosilicate, yttrium disilicate and a combination of both with a 5 vol.% of SiC nanoparticles, after exposure to temperatures ranging between 1000 °C and 1300 °C for 1 – 24 h in air. Their results provide a better understanding of the self-healing process, accounting for the study of this effect at different temperatures and exposure times. The authors also remark the importance of ion diffusion as an explanation for the crack-healing, mechanism not previously considered and discussed. Despite the promising results reported, some considerations should be taken into account. First, as mentioned by the authors, the consumption of both the SiC nanofillers and the YbMS phase limits the self-healing capabilities to a single annealing process. After that, the presence of both SiC and YbMS would not be high enough to trigger the self-healing mechanism when a crack appears. Secondly, the experiments were carried out using sintered bodies, which has already been pointed out in this work that is a less than ideal representation of EBC deposited with techniques currently favoured by the industry. The reduced porosity of the sintered body, for instance, causes that the majority of the reaction between the SiC nanofillers and the oxygen takes place at the cracks.

Thermal sprayed coatings will present higher levels of porosity, which might provide additional reaction centres. This increased formation of YbDS, and the associated volume expansion, might cause excessive compressive levels in the coating, leading to failure [101]. Secondly, the chosen annealing temperature of 1250 °C provides a good self-healing effect; however, is lower than the service temperature at which EBCs are expected to operate (around 1500 °C). It is worth considering if the increased temperature would still allow the described self-healing mechanism to take place as described, or if the kinetics would be altered. Regarding the consideration of kinetics, the addition of steam to the testing atmosphere would be a necessary following step. As it was reported by Opila *et al.* [5], $SiO_2$ will react with steam at temperatures as low as 1200 °C to form gaseous $Si(OH)_4$. If the kinetics of this reaction at high temperatures are higher than the reaction of the amorphous $SiO_2$ with the YbMS to form YbDS and heal the cracks, the self-healing mechanism might be effectively suppressed.

Following the approach of incorporating additives to EBC compositions with a proven performance, Lee [102] produced YbDS APS deposited coatings with the addition of mullite, $Al_2O_3$, $Y_3Al_5O_{12}$ (YAG) or $TiO_2$ (with content below 5 wt.% in all the cases) to reduce the growth rate of the thermally grown oxide (TGO) at the YbDS/Si interface. The growth rate of the TGO has been linked to the failure of EBC systems [69,103,104], making a composition that would reduce its severity very attractive. It was found that the addition of $Al_2O_3$ or $Al_2O_3$-containing compounds reduced the thickness of the TGO up to ~80% when compared to the non-modified YbDS/Si baseline following steam cyclic testing (1 atm pressure, flow velocity of 10 cm/s, 90% $H_2O$/10% $O_2$ environment, 60 min at 1316 °C and 20 min at <100 °C). Although the author remarks the lack of experimental data to fully determine the nature of this phenomenon, it is theorised that the additives produce a beneficial effect not by modifying the oxidiser permeability of the YbDS. On the contrary, a modification of the $SiO_2$ network within the TGO itself, effectively hindering the access of oxidisers to the Si bond layer underneath would explain the reduction of the TGO growth rate. Since the experiments were conducted on APS deposited coatings, having as a baseline a non-modified EBC system that has been extensively proven, this work provides a new interesting trail to follow in the development of the next generation of EBCs. Additionally, given the low concentration of the new additions it would be expected that the impact on the properties of the EBC are somewhat small, as proven by the steam cyclic test, and therefore this could be a reliable and easy approach to improve the operational life of EBCs.

Notable is the work of Turcer *et al.* [36] regarding the exploration of what has been named thermal environmental barrier coatings (TEBC), marrying the concept of thermal insulation in TBCs and protection against the environment in EBCs. To that end, the main four requirements for the development of TEBCs were established: gas-tight protection achieved through a good CTE match with the substrate, high temperature capability or phase stability, low thermal conductivity and resistance against CMAS attack. Experimental and theoretical measurements were performed to allow for a thorough screening of the best potential candidate to be used as a TEBC. CTE match and high temperature phase stability were chosen as the first criterion to be used, allowing for the selection of β-$Yb_2Si_2O_7$, β-$Sc_2Si_2O_7$ and β-$Lu_2Si_2O_7$ as the initial rare earth silicates to be considered. The capability of forming complete solid solutions with the desired rare earth doping elements (i.e. Y, Yb, Sc, Lu) was also considered, before studying the thermal conductivity and CMAS resistance of the potential TEBCs. Extensive theoretical calculations were performed in order to determine the most beneficial combination of rare earth silicates and solid solution, along with the appropriate doping level. The authors concluded that solid solutions alloys of $Y_xYb_{(2-x)}Si_2O_7$ would comply with the requirements identified while improving the current thermal capabilities of EBCs. Further research validating the theoretical calculations for the compositions described would open a new line of research with great potential benefits in the field of environmental barrier coatings.

## 5. Concluding remarks

The introduction of SiC/SiC CMCs in the hot section of gas turbine engines is expected to bring a new revolution to the fields of aerospace and land-based energy power generation in terms of increased efficiency and reduced pollutant and $CO_2$ emission. Before the widespread replacement of the current generation of Ni-based super-alloys can take place, a reliable solution for the corrosion of SiC-based CMCs due to steam and molten salt needs to be introduced. EBCs are presented as the solution to these challenges, and considerable effort has been put into the matter over the last decades. The key element of a successful EBC is a gas-tight morphology of the top coat, preventing the ingress of oxidisers (such as steam) to the underlaying structures. To achieve this, first the appearance of cracks must be avoided, and research has pointed out thermal stresses due to CTE mismatch as the primary origin of such cracks. Nevertheless, this requirement alone is not enough, as proven by the use of mullite or rare earth silicates with multiple polymorphs in the early iterations. Phase transformations at high temperature will induce failure of the coating, so high temperature phase stability is also required. Further research into more complex formulations, such as BSAS, produced promising results in terms of CTE match and phase stability; however, it was found that at temperatures above ~1300 °C the BSAS would react with the $SiO_2$ of the thermally grown oxide layer, producing a glassy sub-product that produced the premature failure of the EBC. Since EBCs are expected to operate at temperatures greatly above 1300 °C, BSAS was discarded due to lack of chemical compatibility between the different layers. Finally, extensive research has been conducted to determine the silica activity and CMAS reactivity of different rare earth silicates candidates, aiming to fulfil the fourth and final requirement of a successful EBC, the effective protection of the substrate against the service environment.

Promising advancements have been made in relation to the performance of these compositions under steam, being now the relative volatilisation rate and fundamental mechanisms understood. The situation is not as clear in regard to molten alkali salts (modelled through the use of CMAS). The deposition technique chosen for the manufacturing of the EBC also plays an essential role, as the physical and chemical properties will be affected. Parameters such as phase content and level of porosity present within the coating play a critical role in the interaction. Although currently APS has been the preferred deposition technique, different thermal spraying techniques, such as HVOF, or novel feedstock presentations such as suspension, could provide further customisation of the characteristics of the coating. Finally, more advanced EBC compositions are already being developed and researched, aiming to open the route towards the next generation. Some noteworthy examples are the addition of SiC nanofillers to ytterbium disilicate coatings in order to provide the system with self-healing capabilities, the addition of $Al_2O_3$ or $Al_2O_3$-containing compounds to modify the oxidation rate of the thermally grown oxide layer, or the introduction of rare earth dopants in solid solution to improve the thermal conductivity of the EBC. Despite the vast amount of research over the last decades, further work is needed to fully understand the corrosion mechanism present in the more promising rare earth silicate candidates.

In conclusion, EBCs represent a fast-paced field with new approaches constantly being researched and reported, aiming to facilitate the transition to a new generation of gas turbines. Due to the nature of the expected applications, research must be conducted in close collaboration with the industry, in order to set realistic deposition and testing standards that closely represent the current manufacturing capabilities of the interested parties as well as the expected service conditions. The exciting milestones achieved in the past few years present a bright picture for this field, projecting an increasing interest and service-ready EBC solutions in the coming years.

## 6. Acknowledgments

This work was supported by the Engineering and Physical Sciences Research Council (grant number EP/L016203/1). The authors would like to thank Dr Acacio Romero Rincon for his helpful discussion and corrections of the manuscript.